\documentclass[a4paper,11pt]{article}
\usepackage{jheppub}

\usepackage{ifpdf}

\ifpdf
\usepackage[dvipsnames]{xcolor}
\else
\usepackage[dvips,dvipsnames]{xcolor}
\fi

\usepackage[utf8]{inputenc}
\usepackage{{amsmath,amsfonts,latexsym, amstext, amssymb, amsthm}}
\usepackage{xspace} 
\usepackage{slashed}
\usepackage[bf]{caption}
\usepackage{subcaption}
\usepackage{comment}
\usepackage{dsfont}
\usepackage{placeins}

\usepackage{breqn}

\usepackage{feynmp}
\newcommand{\cRem}[6]{(\vec{c}_i){}_{#1,#2,#3}^{#4,#5,#6}}

\newcommand{\dd}{\mathrm{d}}

\newcommand{\col}{\;,}
\newcommand{\pnt}{\;.}

\usepackage{feynmp}
\ifpdf
\else
\fi

\usepackage{xkeyval}
\input{picturemacro}

\usepackage{tikz}
\usetikzlibrary{shapes}

\newlength{\vacuumradius}
\newlength{\onshellradius}
\tikzstyle{db}=[circle, black, fill=black, minimum width=\onshellradius, draw, inner sep=0pt]
\tikzstyle{dw}=[circle, black, fill=white, minimum width=\onshellradius, draw, inner sep=0pt]
\tikzstyle{dvac}=[circle, black, fill=lightgray, minimum width=\vacuumradius, inner sep=0pt]

\def\blobhdist{0.4}
\def\blobvdist{0.1}
\def\extradist{0.12}
\def\blobheight{0.7}
\def\blobwidth{\blobhdist*1.5}

\tikzstyle{twoblob}=[ellipse, black, fill=white, minimum width=\blobwidth cm, minimum height=\blobheight cm, draw, inner sep=0pt]
\tikzstyle{threeblob}=[ellipse, black, fill=white, minimum width=\blobhdist cm + \blobwidth cm, minimum height=\blobheight cm, draw, inner sep=0pt]

\newcommand{\bloblabelsize}{\scriptsize}

\newcommand{\drawvlineblob}[2]{
        \draw (#1*\blobhdist-1*\blobhdist,0) -- (#1*\blobhdist-1*\blobhdist,#2*\blobvdist +#2*\blobheight +\blobvdist);
        \draw[white] (#1*\blobhdist-1*\blobhdist-\extradist,0) -- (#1*\blobhdist-1*\blobhdist +\extradist,0);}
\newcommand{\drawtwoblob}[3]{
\node[twoblob] at (#1*\blobhdist-1*\blobhdist+0.5*\blobhdist,#2*\blobvdist +#2*\blobheight-0.5*\blobheight) {\bloblabelsize #3};
}        
\newcommand{\drawthreeblob}[3]{
\node[threeblob] at (#1*\blobhdist-1*\blobhdist +1*\blobhdist,#2*\blobvdist +#2*\blobheight-0.5*\blobheight) {\bloblabelsize #3};
}

\newcommand{\lambdat}{\tilde{\lambda}}

\DeclareMathOperator{\idm}{\mathds{1}}

\newcommand{\YM}{{\mathrm{\scriptscriptstyle YM}}}

\DeclareMathOperator{\tr}{tr}

\DeclareMathOperator{\phaneq}{\phantom{{}=}}

\newcommand{\e}{\operatorname{e}}

\newcommand{\eqndot}{\, .}
\newcommand{\eqncom}{\, ,}
\newcommand{\eqnsem}{\, ;}

\newcommand{\de}{\operatorname{d}\!}

\newcommand{\secref}[1]{section~\ref{#1}}

\newcommand{\cA}{\mathcal{A}}

\newcommand{\cF}{\mathcal{F}}

\newcommand{\cI}{\mathcal{I}}

\newcommand{\cN}{\mathcal{N}}
\newcommand{\cO}{\mathcal{O}}

\newcommand{\cZ}{\mathcal{Z}}

\newcommand{\Nfour}{$\mathcal{N}=4$\xspace}

\newcommand{\NfSYMt}{\Nfour SYM theory\xspace}

\newcommand{\cder}{\D}

\newlength{\eqoff}

\DeclareMathOperator{\D}{D}

\newcommand{\DOT}[1]{\dot{#1}}

\newcommand{\XYsize}{\scriptscriptstyle}

\newcommand{\Rem}{R^{(2)}}
\newcommand{\rem}{R^{(2)}}

\newcommand{\remif}[2]{(R^{(2)}_i)_{\XYsize #1}^{\XYsize #2}}

\newcommand{\interaction}{I}
\newcommand{\Interaction}{\cI}
\newcommand{\interactionr}{\underline{\interaction}}
\newcommand{\Interactionr}{\underline{\Interaction}}
\newcommand{\inttwo}[1][]{\interaction^{(2)}_{#1}}

\newcommand{\intone}[1][]{\interaction^{(1)}_{#1}}

\newcommand{\inttwoi}{\interaction^{(2)}_i}
\newcommand{\inttwoip}{\interaction^{(2)}_{i+1}}
\newcommand{\intonei}{\interaction^{(1)}_i}
\newcommand{\intoneip}{\interaction^{(1)}_{i+1}}

\newcommand{\intoneir}{\interactionr^{(1)}_i}
\newcommand{\intoneipr}{\interactionr^{(1)}_{i+1}}

\newcommand{\intoneif}[2]{(\intonei)_{\XYsize #1}^{\XYsize #2}}

\newcommand{\zone}[1][]{\cZ^{(1)}_{#1}}
\newcommand{\zonei}{\zone[i]}
\newcommand{\zoneip}{\zone[i+1]}
\newcommand{\ztwo}{\cZ^{(2)}}
\newcommand{\ztwoi}{\cZ^{(2)}_i}

\newcommand{\zoneif}[2]{(\zonei)_{\XYsize #1}^{\XYsize #2}}

\newcommand{\dila}{\mathfrak{D}}

\newcommand{\dilatwo}[1][]{\dila^{(2)}_{#1}}

\newcommand{\dilaone}[1][]{\dila^{(1)}_{#1}}
\newcommand{\dilaonei}{\dila^{(1)}_i}
\newcommand{\dilaoneif}[2]{(\dilaonei)_{\XYsize #1}^{\XYsize #2}}

\newcommand{\peps}{\varepsilon}

\newcommand{\sfrac}[2]{{\textstyle\frac{#1}{#2}}}

\newcommand{\amp}{\cA}
\newcommand{\ampco}{\amp}
\newcommand{\ff}{\cF}
\newcommand{\ffco}{\ff}

\numberwithin{equation}{section}

\makeatletter
\DeclareRobustCommand*{\bfseries}{%
  \not@math@alphabet\bfseries\mathbf
  \fontseries\bfdefault\selectfont
  \boldmath
}

\makeatother

\usepackage{lipsum}
\makeatletter
\if@titlepage
  \renewenvironment{abstract}{%
      \titlepage
      \null\vfil
      \@beginparpenalty\@lowpenalty
      \begin{center}%
        \bfseries \abstractname
        \@endparpenalty\@M
      \end{center}}%
     {\par\vfil\null\endtitlepage}
\else
  \renewenvironment{abstract}{%
      \if@twocolumn
        \section*{\abstractname}%
      \else
        \small
        \begin{center}%
          {\bfseries \abstractname\vspace{-.5em}\vspace{\z@}}%
        \end{center}%
        \quotation
      \fi}
      {\if@twocolumn\else\endquotation\fi}
\fi
\makeatother

\newcommand{\beq}{\begin{equation}}
\newcommand{\eeq}{\end{equation}}
\newcommand{\beqa}{\begin{eqnarray}}
\newcommand{\eeqa}{\end{eqnarray}}
\newcommand{\bea}{\begin{aligned}}
\newcommand{\eea}{\end{aligned}}

\allowdisplaybreaks

\title{Two-Loop SL(2) Form Factors and Maximal Transcendentality}
\author{Florian Loebbert, Christoph Sieg, Matthias Wilhelm, Gang Yang}

\begin{document}
\begin{fmffile}{diagramssltwopaper}
\fmfcmd{%
thin := 1pt; 
thick := 2thin;
arrow_len := 3mm;
arrow_ang := 15;
curly_len := 3mm;
dash_len :=0.3; 
dot_len := 0.75mm; 
wiggly_len := 2mm; 
wiggly_slope := 60;
zigzag_len := 2mm;
zigzag_width := 2thick;
decor_size := 5mm;
dot_size := 2thick;
}
\fmfcmd{%
marksize=7mm;
def draw_cut(expr p,a) =
  begingroup
    save t,tip,dma,dmb; pair tip,dma,dmb;
    t=arctime a of p;
    tip =marksize*unitvector direction t of p;
    dma =marksize*unitvector direction t of p rotated -90;
    dmb =marksize*unitvector direction t of p rotated 90;
    linejoin:=beveled;
    drawoptions(dashed dashpattern(on 3bp off 3bp on 3bp));
    draw ((-.5dma.. -.5dmb) shifted point t of p);
    drawoptions();
  endgroup
enddef;
style_def phantom_cut expr p =
    save amid;
    amid=.5*arclength p;
    draw_cut(p, amid);
    draw p;
enddef;
}
\fmfcmd{%
smallmarksize=4mm;
def draw_smallcut(expr p,a) =
  begingroup
    save t,tip,dma,dmb; pair tip,dma,dmb;
    t=arctime a of p;
    tip =smallmarksize*unitvector direction t of p;
    dma =smallmarksize*unitvector direction t of p rotated -90;
    dmb =smallmarksize*unitvector direction t of p rotated 90;
    linejoin:=beveled;
    drawoptions(dashed dashpattern(on 2bp off 2bp on 2bp) withcolor red);
    draw ((-.5dma.. -.5dmb) shifted point t of p);
    drawoptions();
  endgroup
enddef;
style_def phantom_smallcut expr p =
    save amid;
    amid=.5*arclength p;
    draw_smallcut(p, amid);
    draw p;
enddef;
}

\begingroup\parindent0pt
\begin{flushright}\footnotesize
\texttt{HU-MATH-2016-17}\\
\texttt{HU-EP-16/31} 
\end{flushright}
\vspace*{4em}
\centering
\begingroup\Large
\bf

Two-Loop SL(2) Form Factors and Maximal Transcendentality 

\par\endgroup
\vspace{2.5em}
\begingroup\normalsize\bf
Florian Loebbert$^{\text{a}}$,  Christoph Sieg$^{\text{a,b}}$, \\ Matthias Wilhelm$^{\text{a,b,c}}$, Gang Yang$^{\text{d,a}}$
\par\endgroup
\vspace{1em}
\begingroup\itshape
$^{\text{a}}$Institut für Physik, Humboldt-Universität zu Berlin,\\
Zum Großen Windkanal 6, 12489 Berlin, Germany\\
$^{\text{b}}$Institut für Mathematik, Humboldt-Universität zu Berlin,\\
Zum Großen Windkanal 6, 12489 Berlin, Germany\\
$^{\text{c}}$Niels Bohr Institute, Copenhagen University,\\
Blegdamsvej 17, 2100 Copenhagen \O{}, Denmark\\
$^{\text{d}}$CAS Key Laboratory of Theoretical Physics, Institute of Theoretical Physics,\\
Chinese Academy of Sciences, Beijing 100190, China
\par\endgroup
\vspace{1em}
\begingroup\ttfamily
loebbert@physik.hu-berlin.de, csieg@physik.hu-berlin.de, matthias.wilhelm@nbi.ku.dk, yangg@itp.ac.cn \\
\par\endgroup
\vspace{2.5em}

\vspace{2.5em}
\endgroup

\begin{abstract}
\noindent
Form factors of composite operators in the SL(2) sector of \NfSYMt are studied up to two loops via the on-shell unitarity method. The non-compactness of this subsector implies the novel feature and technical challenge of an unlimited number of loop momenta in the integrand's numerator.
At one loop, we derive the full minimal form factor to all orders in the dimensional regularisation parameter. 
At two loops, we construct the complete integrand for composite operators with an arbitrary number of covariant derivatives, and we obtain the remainder functions as well as the dilatation operator for composite operators with up to three covariant derivatives. 
The remainder functions reveal curious patterns suggesting a 
hidden maximal uniform transcendentality for the full form factor.
Finally, we speculate about an extension of these patterns to QCD.

\end{abstract}

\thispagestyle{empty}

\newpage

\hrule
\tableofcontents
\afterTocSpace
\hrule
\afterTocRuleSpace


\section{Introduction and summary}
\label{sec:intro}

Recent years have seen increasing interest in the study of form factors. 
In particular, $\cN=4$ super Yang--Mills (SYM) theory in four spacetime dimensions furnishes a rich setting for studying their properties.
Form factors in ${\cal N}=4$ SYM theory were initially investigated in 1985~\cite{vanNeerven:1985ja}. 
Only recently, this subject was reconsidered, first at strong coupling \cite{Alday:2007he, Maldacena:2010kp} and then at weak coupling \cite{Brandhuber:2010ad, Bork:2010wf},
and has subsequently aroused a lot more attention 
\cite{Brandhuber:2011tv,Bork:2011cj,Henn:2011by,Gehrmann:2011xn,Brandhuber:2012vm,Bork:2012tt,Engelund:2012re,Johansson:2012zv,Boels:2012ew, Gao:2013dza, Penante:2014sza, Brandhuber:2014ica, Bork:2014eqa, Wilhelm:2014qua, Nandan:2014oga, Loebbert:2015ova, Frassek:2015rka, Boels:2015yna, Huang:2016bmv, Koster:2016ebi, Koster:2016loo, Chicherin:2016qsf, Brandhuber:2016fni,  Bork:2016hst, Bork:2016xfn, He:2016dol, Caron-Huot:2016cwu, Brandhuber:2016xue, He:2016jdg,Yang:2016ear,Ahmed:2016vgl}. 
For recent reviews, see also the theses \cite{Wilhelm:2016izi, Penante:2016ycx}.

An $n$-point form factor describes the interaction of $n$ on-shell particles with momenta $p_i$ ($p_i^2=0$) and a gauge-invariant local composite operator ${\cal O}(x)$. Translated into momentum space, it takes the form
\begin{equation}
\label{eq:formfactor}
{\cal F}_{\mathcal{O}}(1,...,n;q) = \int \de^D x \e^{-i q \cdot x} \langle 1\cdots n|{\cal O}(x) | 0\rangle\eqncom
\end{equation}
where in this paper all form factors are considered to be colour-ordered super form factors and $q$ is off-shell ($q^2\neq0$).
The motivation for studying form factors is multifold. Firstly, given the tremendous progress on on-shell techniques for scattering amplitudes (see e.g.\ \cite{Elvang:2013cua, Dixon:2013uaa, Henn:2014yza}), it is desirable to transfer these powerful methods to the computation of more general quantities. Form factors combine external on-shell and off-shell 
objects and are thus ideally suited for this purpose. 
Showing a rich structure, they also provide new testing grounds for the further development of on-shell techniques. 
Secondly, forming a bridge between on-shell and off-shell observables, the study of form factors should yield novel insights into the hidden symmetries of the underlying theory. Moreover, we should emphasise that form factors are interesting in their own right.  Sudakov form factors, for instance, have played a key role in the understanding of infrared singularities in gauge theories \cite{Mueller:1979ih, Collins:1980ih, Sen:1981sd, Magnea:1990zb}. Furthermore, they are closely related to Higgs production via gluon fusion as well as to Drell-Yan production in the standard model \cite{Georgi:1977gs, Drell:1970wh}.

The study of form factors of non-protected operators, as well as their connection to the spectral problem of planar ${\cal N}=4$ SYM theory, was recently pushed forward in \cite{Wilhelm:2014qua, Nandan:2014oga, Loebbert:2015ova, Brandhuber:2016fni,Ahmed:2016vgl,Caron-Huot:2016cwu}.
In particular, the form factor remainder was made a central concept and it was shown how to extract the dilatation operator from form factor data.
While previous studies mostly focused on 
individual operators or compact sectors, in this paper we consider operators in the non-compact SL(2) sector. 
The crucial new feature of non-compactness
brings technical challenges and reveals novel properties.

Operators in the SL(2) sector of $\mathcal{N}=4$ SYM theory involve a single type of complex scalar $X$ and an arbitrary number of insertions of covariant derivatives $\cder^\mu$ projected onto a lightlike vector $\tau^\mu$:
\begin{align}\label{eq:defstates}
\cO_{\vec n}=\tr \bigg[ \prod_{j=1}^L\frac{(\cder^{+})^{n_j}}{n_j!} X \bigg] \,, \qquad
\cder^{+} = \cder^\mu \tau_\mu \pnt
\end{align}
The length of the operator, i.e.\ the number of scalars $X$, is denoted by $L$.
The configuration of a certain state is thus completely specified by the vector
\begin{equation}
\label{eq:spin-config}
{\vec n} = (n_1,\dots,n_L) \col
\end{equation}
which is identified with its images under cyclic permutations. 
We refer to the total number of covariant derivatives, $M=\sum_{j=1}^Ln_j$, as the total spin or  the total magnon number.%
\footnote{Note that the term magnon usually refers to a spin-wave. We use this notation due to the strong connection to spin chains and integrability, see \cite{Beisert:2010jr} for a review.}
In what follows, we will also speak of form factors in the SL(2) sector, meaning form factors with a composite operator from the above SL(2) sector.
Throughout this paper, we will mostly focus on minimal form factors, i.e.\ on the case where we have as many external fields $n$ as fields in the operator $L$.%
\footnote{We only consider $L\geq3$ here. For $L=2$, the remainder of the minimal form factor is a pure number as all dependence on the kinematics factors out.}
Moreover, we will restrict ourselves to the planar limit; the employed methods are also applicable for finite $N$ though. 

At the technical level, the unlimited number of covariant derivatives in the SL(2) sector implies that the form factor integrand at fixed loop order involves arbitrarily high powers of loop momenta in the numerator. This is a completely new feature as compared to amplitude computations or to the previous studies of form factors with operators in compact sectors. As demonstrated in this paper, on-shell methods can still be efficiently applied in this case. 

At one-loop order, we derive the full integrated form factor for all composite operators in the SL(2) sector. At two loops, we obtain the full integrand for an arbitrary magnon number using the unitarity method \cite{Bern:1994zx,Bern:1994cg, Britto:2004nc}. Solving the resulting integrals via traditional methods becomes more and more challenging for increasing magnon number. In this paper, we present the complete integrated result up to total spin three. We leave the solution for an arbitrary magnon number for future work. Besides universal IR divergences, the form factors contain UV divergences which allow us to extract 
the dilatation operator of the theory;
our results (up to three magnons) 
match perfectly with the previously obtained data for the dilatation operator given in \cite{Eden:2006rx,Belitsky:2006av, Zwiebel:2008gr}.
Despite the relatively small magnon numbers, we find very interesting patterns for the two-loop remainder function as well as for the finite terms at one-loop order. These observations indicate that form factors in ${\cal N}=4$ SYM theory have 
`hidden' uniform maximal transcendentality
 for any operator, as will now be explained.

The first pattern to be noted is that the remainder function contains a universal piece of maximal transcendentality four, which is identical to the BPS result \cite{Brandhuber:2014ica}. 
This was first observed in the compact SU(2) sector and conjectured to be a generic feature of the theory in \cite{Loebbert:2015ova}. So far, further support for this conjecture was already found in the SU(2$|$3) sector studied in \cite{Brandhuber:2016fni}. 
In the SL(2) sector, also lower transcendental functions occur for remainders of non-protected operators, i.e.\ combinations of polylogarithms and constants such as $\zeta_3$ and $\pi^2$, whose total degree does not add up to four.
While this is expected from the SU(2) and SU(2$|$3) sectors \cite{Loebbert:2015ova,Brandhuber:2016fni},
a curious
observation of the present paper is that the functions of lower transcendentality 
come with coefficients composed of (generalised) harmonic numbers or factors of $\frac{1}{m_i-n_i}$, where $m_i, n_i$ are magnon numbers as in \eqref{eq:spin-config}. 
In particular, if also the latter are assigned an appropriate degree,
the two-loop remainder is of uniform transcendentality four; 
similar structures can be seen in the one-loop form factor results. We refer to this assignment as hidden transcendentality.%
\footnote{%
Note that the assignment of transcendentality to harmonic numbers is standard in the context of anomalous dimensions, see e.g.\ \cite{Kotikov:2002ab}. 
An assignment of transcendentality to similar fractions occurred in different contexts in the literature, see e.g.\ \cite{Fleischer:1997bw,Fleischer:1998nb,Bianchi:2013sta}.
}

In order to illustrate the above idea, let us give the following list of exemplary expressions with a (formal) degree of transcendentality $k$:
\begin{center}
\label{eq:trans-k-example}
\setlength\tabcolsep{3mm}
\begin{tabular}{|@{\hspace{2mm}}c|c|c|c|c|c|c|c|c|}\hline
$\frac{1}{\peps^k}$&$\pi^k$&$\zeta_k$&$\text{Li}_k(u_i)$&$\log^k(u_i)$&$S_{n_j}^{(k)}$&$\frac{1}{(m_j-n_j)^k}$&$\sum\limits_{j=1}^{k}\frac{1}{j}\frac{(1-u_i)^{j}}{v_i^j}$&$\frac{1}{(m_j-n_j)^k}\frac{u_i^{k}}{v_i^{k}}$\,
\\\hline
\end{tabular}
\end{center}
Here, 
$u_i$ and $v_i$ represent kinematical variables, $m_i, n_i$ are magnon numbers  as in \eqref{eq:spin-config}, and the dimensional regularisation parameter $\peps$ is assigned a transcendentality degree $-1$, as normally done in the literature.
The property of hidden uniform transcendentality can already be observed for the explicit expression for the two-loop dilatation operator given in \cite{Belitsky:2006av}. The main
observation made here is that this property appears to hold for the full form factor including functions with kinematical dependence, such as the remainder function.
Note that in the case of the latter, we allow for generalisations of harmonic numbers or inverse powers of magnon numbers by inclusion of kinematical variables, such as $\sum_{j=1}^{k}\frac{1}{j}\frac{(1-u_i)^{j}}{v_i^j}$ in the list above.
A more detailed explanation as well as explicit examples will be given in section \ref{sec:remainderstwoloop}.

The above leads us to the following conjecture:
\emph{The full form factor for any given operator in ${\cal N}=4$ SYM theory has hidden  uniform transcendentality. }
 
Note that, in sectors beyond SL(2), the notion of magnon number might need to be suitably generalised, e.g.\ to the total number of excitations  in the spin-chain picture for composite operators.
Further note that the property conjectured above is not visible from the study of operators in compact subsectors as performed in \cite{Nandan:2014oga,Loebbert:2015ova, Brandhuber:2016fni}, unless they are embedded into a larger non-compact sector (i.e.\ through inclusion of an arbitrary number of covariant derivatives).

Further backing for the above conjecture comes from the relation between SL(2) operators and 
lightlike Wilson loops. It is known that twist-two operators (i.e.\ length-two operators in the SL(2) sector) in the large spin limit are related to lightlike cusped Wilson loops \cite{Korchemsky:1988si, Korchemsky:1992xv}, which have uniform transcendentality \cite{BES06}. This is related to the fact that harmonic numbers $S_m^{(l)}$ become transcendental numbers of degree $l$ for $m\rightarrow \infty$. 
Similarly, the generalisations of harmonic numbers become (poly)logarithms; e.g.\ $\sum_{j=1}^{m}\frac{1}{j}\frac{(1-u_i)^{j}}{v_i^j}\to-\log\big(1-\frac{1-u_i}{v_i}\big)$ for $m\to\infty$. 
Generic SL(2) operators are related to multi-edge lightlike Wilson loops. Since lightlike Wilson loops, which are dual to amplitudes, have uniform transcendentality in ${\cal N}=4$ SYM theory \cite{Drummond:2008aq, Bern:2008ap, DelDuca:2010zg, ArkaniHamed:2012nw}, this provides more support for our conjecture.

While the theory underlying the considerations of this paper is $\mathcal{N}=4$ SYM theory, we expect that the above insights may also teach us interesting lessons about similar calculations in QCD. In particular, one may wonder whether the maximal transcendentality principle \cite{Kotikov:2002ab,Kotikov:2004er} extends to the notion of hidden transcendentality indicated above, see also \cite{Brandhuber:2012vm}. We will further discuss this point in the outlook in section \ref{sec:conclusion}.

The paper is organised as follows. In section \ref{sec:tree}, we give the tree-level form factors needed for the unitarity computations. Then, the full one-loop minimal form factor is obtained in section \ref{sec:one-loop} and the one-loop dilatation operator and finite terms are extracted. In section \ref{sec:two-loop}, we construct the complete two-loop integrand and obtain the two-loop dilatation operator and the remainder functions for operators up to total magnon number three. We moreover discuss the interesting pattern and transcendentality property for the remainder functions. A conclusion and outlook is given in section \ref{sec:conclusion}. 
In appendix \ref{app: one-loop integral}, we give technical details on evaluating the one-loop form factor integrals. 
The complete remainder functions up to total magnon number three are given in terms of their densities in appendix \ref{app: remainder densities}.


\section{Tree level}
\label{sec:tree}

In this section, we provide the tree-level form factors in the SL(2) sector which occur in the unitarity cuts in the subsequent sections.

\paragraph{Minimal form factors.}

Within the SL(2) sector, a \emph{minimal} form factor for a single-trace operator of length $L$ contains $L$ external on-shell scalar states. While the state can be conveniently expressed in terms of oscillators acting on a (spin chain) vacuum \cite{Beisert:2003jj}, the minimal form factor is simply obtained by replacing these oscillators by spinor helicity variables $(\lambda,\tilde{\lambda})$ and the SU(4) Gra{\ss}mann spinors $\eta^A$ \cite{Wilhelm:2014qua}, see also \cite{Beisert:2010jq,Zwiebel:2011bx}. 
For the SL(2) sector, each scalar $X=\phi_{34}$ at position $i$ contributes a factor $\eta^3_i \eta^4_i$ and each covariant derivative $\cder^+$ contributes a factor $p^+_i=p_i^\mu \tau_\mu$.%
\footnote{Note that we are suppressing a factor of the imaginary unit $i$ from the (covariant) derivative.}
For explicitness, but without loss of generality, we choose 
$\tau_\mu = (\sigma_\mu)^{1\DOT1}$, such that $\cder^{+} = \cder^{1\DOT1}$ and $p^+=p^{1\DOT1}=\lambda^1\tilde{\lambda}^{\DOT1}$,
 when required.
The minimal form factor can thus be written as
\begin{equation}
\label{eq:tree-minimal}
\ffco^{(0)}_{\cO_{\vec n}}(1,2,\dots,L;q)=\delta^4\left(\sum_{i=1}^L p_i-q\right)
\sum_{k=1}^L\left(\prod_{j=1}^L\frac{(p_{k+j}^+)^{n_j}}{n_j!}\eta_{k+j}^3\eta_{k+j}^4\right) \,,
\end{equation}
where the sum over $k$ is due to the cyclicity of the colour-ordered form factor, which in the minimal case is essentially given by the manifestly cyclically invariant vertex of the operator.
Having fixed the type of scalar $X$, a given operator or minimal form factor, respectively, is hence characterised by the derivative configuration $\vec{n}=(n_1,\dots,n_L)$.


\paragraph{Next-to-minimal form factors.}

As important building blocks for the two-loop computation via unitarity, we will also need the next-to-minimal tree-level form factors, which have $L+1$ external states. These can be easily computed via Feynman diagrams. 
The results coincide with the corresponding tree-level expressions for general operators obtained via the twistor action in \cite{Koster:2016loo}.

At MHV level, there are two different configurations of external states to be considered for the next-to-minimal form factors. 
The first configuration consists of $L-1$ scalar external states and two fermionic external states. The corresponding next-to-minimal form factors can be obtained from the minimal form factors by replacing $X_i=\phi_i^{34}\to \bar{\psi}^3_i\bar{\psi}^4_{i+1}-\bar{\psi}^4_i\bar{\psi}^3_{i+1}$ in the external states of the minimal form factor \eqref{eq:tree-minimal} for each $i=1,\dots, L$ according to 
\begin{equation}
\label{eq: non-minimal replacement rule 1}
 \frac{(p_{i}^+)^{n_j}}{n_j!}\eta_{i}^3\eta_{i}^4\to\frac{1}{\langle i\, i\textrm{+}1\rangle}\sum_{m=0}^{n_j}\frac{(p_{i}^+)^{m}}{m!}\frac{(p_{i+1}^+)^{n_j-m}}{(n_j-m)!}(\eta_i^3\eta_{i+1}^4 - \eta_i^4\eta_{i+1}^3)\col
\end{equation}
where we also have to relabel $p_l^+\to p_{l+1}^+$ for $l=i+1,\dots,L$ and include the momentum $p_{L+1}$ in the momentum-conserving delta function.

The second configuration consists of $L$ scalar external states and one positive-helicity gluon external states. 
For this configuration, two qualitatively different cases occur, which are distinguished via the $\overline{\text{SU}(2)}$ indices, i.e.\ the indices $\dot\alpha$ of $\tilde\lambda$.
In the first case, all $\overline{\text{SU}(2)}$ indices remain on the scalars.
This process can only be attributed to a pair of neighbouring scalars, not to a single scalar, as gauge invariance is achieved by summing two Feynman diagrams: one with the left scalar emitting the gluon to the right and the other with the right scalar emitting the gluon to the left such that the gluon appears 
between the two scalars in the colour-ordered expression.
We can obtain the contribution of this case from the minimal form factors \eqref{eq:tree-minimal} by replacing $X_iX_{i+1}\to X_ig^+_{i+1}X_{i+2}$ in the external states for $i=1,\dots,L$ according to 
\begin{equation}
\label{eq: non-minimal replacement rule 3}
 \frac{(p_{i}^+)^{n_j}}{n_j!}\eta_{i}^3\eta_{i}^4\frac{(p_{i+1}^+)^{n_{j+1}}}{n_{j+1}!}\eta_{i+1}^3\eta_{i+1}^4
 \to\frac{{\langle i\, i\textrm{+}2\rangle}}{{\langle i\, i\textrm{+}1\rangle} {\langle i\textrm{+}1\, i\textrm{+}2\rangle} }\frac{(p_{i}^+)^{n_j}}{n_j!}\frac{(p_{i+2}^+)^{n_{j+1}}}{n_{j+1}!}\eta_i^3\eta_i^4\eta_{i+2}^3\eta_{i+2}^4\col
\end{equation}
where we also have to relabel $p_l^+\to p_{l+1}^+$ for $l=i+2,\dots,L$ and include the momentum $p_{L+1}$ in the momentum-conserving delta function.
In the second case, at least one $\overline{\text{SU}(2)}$ index from scalar $i$ lands on the gluon. 
Two Feynman diagrams contribute to this case: the emission of a gluon by scalar $i$, and the insertion of a polarisation vector into the gauge-field part of the covariant derivative acting on this scalar. 
Their sum is gauge invariant, and, as they are only involving the scalar $i$, we can write this contribution 
by replacing $X_i\to X_ig^+_{i+1}+g^+_iX_{i+1}$ in \eqref{eq:tree-minimal} for $i=1,\dots,L$ according to
\begin{equation}
\label{eq: non-minimal replacement rule 2}
\begin{aligned}
\frac{(p_{i}^+)^{n_j}}{n_j!}\eta_{i}^3\eta_{i}^4\to 
&+\frac{\lambda_i^1\lambdat_{i+1}^{\DOT1}}{\langle i\, i+1\rangle}\sum_{m=1}^{n_j}\frac{1}{m}\frac{(p_{i}^+)^{n_j-m}}{(n_j-m)!}\frac{(p_{i+1}^+)^{m-1}}{(m-1)!}\eta_i^3\eta_{i}^4\\
&+\frac{\lambdat_{i}^{\DOT1}\lambda_{i+1}^1}{\langle i\, i+1\rangle}\sum_{m=1}^{n_j}\frac{1}{m}\frac{(p_{i}^+)^{m-1}}{(m-1)!}\frac{(p_{i+1}^+)^{n_j-m}}{(n_j-m)!}\eta_{i+1}^3\eta_{i+1}^4
\col
\end{aligned}
\end{equation}
where we again relabel $p_l^+\to p_{l+1}^+$ for $l=i+1,\dots,L$ and include the momentum $p_{L+1}$ in the momentum-conserving delta function.

We acquire the next-to-minimal MHV form factor by applying the replacements rules \eqref{eq: non-minimal replacement rule 1}--\eqref{eq: non-minimal replacement rule 2} to all fields in the operator and summing the resulting expressions.
The next-to-minimal NMHV form factor can be obtained from its MHV counterpart via conjugation, as described in \cite{Nandan:2014oga} in our conventions.

\section{One loop}
\label{sec:one-loop}

In this section, we calculate the one-loop correction to the minimal form factors in the SL(2) sector and extract from them the one-loop dilatation operator and the finite part.

\subsection{Unitarity construction}

Let us write the loop expansion of the minimal form factor in the following form:
\begin{equation}
	\label{eq: loop correction}
	\cF_{\cO}= \bigg( 1+ \sum_{\ell=1}^\infty g^{2\ell} \Interaction^{(\ell)} \bigg) \cF_{\cO}^{(0)} \eqncom
\end{equation}
where 
\begin{equation}\label{gdef}
g^2 = \frac{g_\YM^2N_{\text{c}}}{(4\pi)^2}(4\pi\e^{- \gamma_{\text{E}}})^\peps
\end{equation}
is the effective planar coupling constant. 
For operators such as BPS operators or the Konishi primary, which are eigenstates
 under renormalisation, $\Interaction^{(\ell)}$ is simply the ratio of the $\ell$-loop and tree-level form factor.
However, this is not true in general due to operator mixing.
In particular, the loop corrections to vanishing tree-level form factors can be non-vanishing.
It is therefore important to promote $\Interaction^{(\ell)}$ to an {\it operator} that acts on the tree-level form factor $\cF_{\cO}^{(0)}$ and can create 
different tree-level form factors from it. 
Examples in the SU(2) sector and for general operators can be found in \cite{Loebbert:2015ova} and \cite{Wilhelm:2014qua}, respectively.

In the planar limit, connected $\ell$-loop interactions can maximally involve $\ell+1$ neighbouring fields in the colour-ordered form factor at a time. Hence, 
$\Interaction^{(\ell)}$ can be written as a sum over connected interaction densities each of which involves $\ell+1$ adjacent external legs plus disconnected products of lower-loop interaction densities.
At one-loop order, the maximal interaction range is two, and we can write%
\footnote{We use the italic $\interaction$ for the density and the calligraphic $\Interaction$ for its sum.}
\begin{equation}
	\label{eq: one-loop interaction}
	\Interaction^{(1)}=\sum_{i=1}^L \interaction^{(1)}_{i\,i+1} \,,
\end{equation}
where $\interaction^{(1)}_{i\,i+1}$ acts on the external fields $i$ and $i+1$ and cyclic identification $i+L\sim i$ is understood.
$\interaction^{(1)}_{i\,i+1}$ can be given in an operatorial form as
\begin{align}
\label{eq:one-loop-I-operator-form}
\interaction^{(1)}_{i,i+1} = \sum_{n_i,m_i = 0}^{\infty} (\interaction^{(1)}_i)_{n_1, n_2}^{m_1, m_2}  \frac{(p_i^+)^{m_1}}{m_1!} \frac{(p_{i+1}^+)^{m_2}}{m_2!} \bigg( {\partial \over \partial p^+_i} \bigg)_{p_i^+=0}^{n_1} \bigg( {\partial \over \partial p^+_{i+1}} \bigg)_{p_{i+1}^+=0}^{n_2} \,,
\end{align}
where 
$(\interaction_i)_{n_1, n_2}^{m_1, m_2}$ is the matrix element of the interaction density that corresponds to changing the derivative
configuration at momenta $p_i$, $p_{i+1}$ in the tree-level form factor \eqref{eq:tree-minimal} from $n_1$, $n_2$ to $m_1$, $m_2$, respectively.%
\footnote{Note that for compactness we use only the index $i$ instead of $i,i+1$ for the symbol $I$ if the range of the interaction is obvious, e.g.\ due to explicit specification of the spin numbers.}
Here, $\big( {\partial \over \partial p} \big)_{p=0}^{n}$ should be understood as acting to the right such that  
\begin{equation}
\bigg( {\partial \over \partial p} \bigg)_{p=0}^{n} f(p) = {\partial^n f(p) \over \partial p^n} \bigg|_{p=0} \eqncom
\end{equation} 
where the action is understood to be on the polynomial in $p_i^+$, not on the momentum-conserving delta function. 

\begin{figure}[tbp]
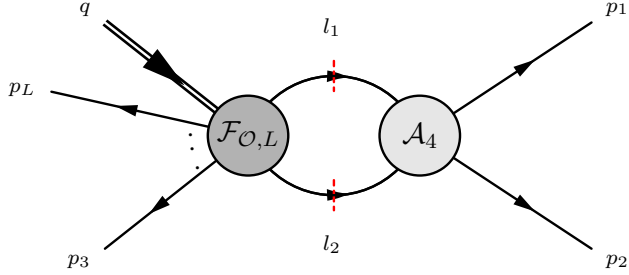

 \centering
$
\settoheight{\eqoff}{$\times$}%
\setlength{\eqoff}{0.5\eqoff}%
\addtolength{\eqoff}{-12.0\unitlength}%
\raisebox{\eqoff}{%
\fmfframe(2,2)(2,2){%
\begin{fmfchar*}(80,30)
\fmfleft{vp3,vp,vpL,vq}
\fmfright{vp2,vp1}
\fmf{dbl_plain_arrow,tension=1.2}{vq,v1}
\fmf{plain_arrow,tension=0}{v1,vpL}
\fmf{plain_arrow,tension=1.2}{v1,vp3}
\fmf{plain_arrow,left=0.7,label=$\scriptstyle l_1\,,$,l.d=15}{v1,v2}
\fmf{plain_arrow,right=0.7,label=$\scriptstyle l_2\,,$,l.d=15}{v1,v2}
\fmf{phantom_smallcut,left=0.7,tension=0}{v1,v2}
\fmf{phantom_smallcut,right=0.7,tension=0}{v1,v2}
\fmf{plain_arrow}{v2,vp1}
\fmf{plain_arrow}{v2,vp2}
\fmfv{decor.shape=circle,decor.filled=30,decor.size=30,label=$\ffco_{\cO,,L}$,label.dist=0}{v1}
\fmfv{decor.shape=circle,decor.filled=10,decor.size=30,label=$\ampco_{4}$,label.dist=0}{v2}
\fmffreeze
 \fmfcmd{pair vertq, vertpone, vertptwo, vertpthree, vertpL, vertone, verttwo; vertone = vloc(__v1); verttwo = vloc(__v2); vertq = vloc(__vq); vertpone = vloc(__vp1); vertptwo = vloc(__vp2); vertpthree = vloc(__vp3);vertpL = vloc(__vpL);}
 \fmfiv{label=$\scriptstyle q$}{vertq}
 \fmfiv{label=$\scriptstyle p_1$}{vertpone}
 \fmfiv{label=$\scriptstyle p_2$}{vertptwo}
 \fmfiv{label=$\scriptstyle p_3$}{vertpthree}
 \fmfiv{label=$\scriptstyle p_L$}{vertpL}
 \fmfiv{label=$\cdot$,l.d=20,l.a=-150}{vertone}
 \fmfiv{label=$\cdot$,l.d=20,l.a=-165}{vertone}
 \fmfiv{label=$\cdot$,l.d=20,l.a=-180}{vertone}
\end{fmfchar*}%
}}%
$
\caption{The double cut of one-loop form factor in the $s_{12}$-channel.}
\label{fig: one-loop double cut}
\end{figure}

\paragraph{One-loop interaction.}
The loop correction $(\interaction_i^{(1)})_{n_1, n_2}^{m_1, m_2}$ can be computed using the unitarity method. We will be comparatively brief here; see e.g.\ \cite{Wilhelm:2014qua,Nandan:2014oga,Loebbert:2015ova,Wilhelm:2016izi} for details of similar calculations and \cite{Loebbert:2015ova} for details on our conventions.
As only interactions between two adjacent fields occur at one-loop order in the planar theory, it is enough to consider the double cut shown in Fig.~\ref{fig: one-loop double cut}. 
The cut integrand takes the form
\begin{align}
\label{eq:one-loop-double-cut}
& \int \de{\rm LIPS}(l_1, l_2) \de^4 \eta_{l_1} \de^4 \eta_{l_2} {\cal F}^{(0)}_{\cal O}(l_1^X, l_2^X, p_3,..,p_L;q) {\cal A}^{(0)}_4(-l_2^{\bar X}, -l_1^{\bar X}, p_1^X, p_2^X) \\*
&= \sum_{{n_i} = 0}^{\infty} 
\bigg[ \int \de{\rm LIPS}(l_1, l_2) {\frac{(l_1^+)^{n_1}}{n_1!}\frac{(l_2^+)^{n_2}}{n_2!}} \frac{\langle l_1 l_2 \rangle \langle 1\,2\rangle}{\langle l_1 1 \rangle \langle l_2 2 \rangle} \bigg]
\bigg( {\partial \over \partial p^+_1} \bigg)_{p_1=0}^{{n_1}} \bigg( {\partial \over \partial p^+_2} \bigg)_{p_2=0}^{{n_2}} {\cal F}^{(0)}_{\cal O}(p_1,..,p_L;q)\,, \nonumber
\end{align}
where the minimal tree-level form factor is given in \eqref{eq:tree-minimal} and the four-point amplitude is given by the standard MHV expression. The two-particle phase-space measure is denoted by dLIPS.

The full integrand can be obtained from the cut integrand by adding the cut propagators and taking the cut momenta $l_{1,2}$ to be off-shell. Explicitly, the square bracket part on the right hand side of \eqref{eq:one-loop-double-cut}, now written for general $i$, lifts to the one-loop integral as
\begin{equation}
\label{eq:one-loop-density-integrand}
\begin{aligned}
(\interaction^{(1)}_i)_{{n_1}, {n_2}} &=-s_{i\,i+1}\FDinline[triangle,momentum,twolabels,labelone=\scriptscriptstyle p_i,labeltwo=\scriptscriptstyle p_{i+1}]\frac{(l_1^+)^{n_1}}{n_1!} \frac{(l_2^+)^{n_2}}{n_2!}\\
&\equiv -s_{i\,i+1}(\e^{\gamma_{\text{E}}}\mu^2)^{2-\frac{D}{2}}\frac{1}{n_1!n_2!}\int\frac{\de^Dl_1}{i\pi^{\frac{D}{2}}}\frac{(l_1^+)^{n_1} (l_2^+)^{n_2}}{l_1^2l_2^2(l_1-p_i)^2}  \col
\end{aligned}
\end{equation}
where $l_2=p_i+p_{i+1}-l_1$. The 't Hooft mass is denoted by $\mu$, the spacetime dimension is $D=4-2\peps$ and $\gamma_{\text{E}}$ is the Euler-Mascheroni constant.
This expression agrees with the one already found in \cite{Wilhelm:2014qua}.

For given small magnon numbers $n_1$ and $n_2$,
it is straightforward to perform  standard Passarino-Veltmann (PV) reduction \cite{Passarino:1978jh} and  integration-by-parts (IBP) reduction \cite{Chetyrkin:1981qh, Tkachov:1981wb}, which reduces the integral to the scalar
bubble integral \eqref{B1}. 
The integral \eqref{eq:one-loop-density-integrand}
evaluates to a polynomial in terms of the lightcone components of the two 
external momenta as
\begin{align}\label{eq:one-loop-integral-expansion} 
(\interaction^{(1)}_i)_{{n_1}, {n_2}} = \sum_{\substack{m_1,m_2=0\\m_1+m_2=n_1+n_2}}^{{n_1}+{n_2}} (\interaction^{(1)}_i)_{{n_1}, {n_2}}^{{m_1}, {m_2}} \frac{(p_i^+)^{{m_1}}}{m_1!} \frac{(p_{i+1}^+)^{{m_2}}}{m_2!} 
 \pnt
\end{align}

As shown in appendix \ref{app: one-loop integral}, the coefficients 
in the above expression can even be evaluated
to a closed form in the generic case:
\begin{equation}
\begin{aligned}\label{eq:one-loop-integral-result} 
(\interaction^{(1)}_i)_{{n_1}, {n_2}}^{{m_1},{m_2}}
&=\frac{ {m_1}!{m_2}!}{\Gamma(\tfrac{D}{2}+{m_1}-1)\Gamma(\tfrac{D}{2}+{m_2}-1)}
\frac{\Gamma(D-2)}{\Gamma(\tfrac{D}{2}-1)}\FDinline[bubble,twolabels,labelone=\scriptscriptstyle p_i,labeltwo=\scriptscriptstyle p_{i+1}]\\
&\phantom{{}={}}
\sum_{j=0}^{\min({n_1},{n_2},{m_1},{m_2})}
\frac{(-1)^j\Gamma(\tfrac{D}{2}+{n_1}+{n_2}-j-1)}{j!({m_1}-j)!({m_2}-j)!}
\\
&\hphantom{{}={}\sum_{j=0}^{\min({n_1},{n_2},{m_1},{m_2})}}
\sum_{r=\max(0,{m_2}-{n_2})}^{\min({m_2},{n_1})-j}\binom{{m_1}-j}{{n_1}-j-r}\binom{{m_2}-j}{r}
\\
&\hphantom{{}={}
\quad\sum_{r=\max(0,{m_2}-{n_2})}^{\min({m_2},{n_1})-j}}
\frac{\Gamma(\tfrac{D}{2}+{n_2}-{m_2}+2r-2)\Gamma(\tfrac{D}{2}+{n_1}+{m_2}-2j-2r-1)}{\Gamma(D+{n_1}+{n_2}-2j-3)}
\col
\end{aligned}
\end{equation}
where the diagram denotes the standard scalar bubble integral given in \eqref{B1}.
We have checked in numerous special cases that the general result agrees with the one obtained using PV and IBP reduction.

The result \eqref{eq:one-loop-integral-result} contains the full $\peps$-expansion, which, for example, also includes the rational contribution.\footnote{The cut-constructible part has been computed in  \cite{Wilhelm:2014qua}.}  This will be important for computing the two-loop dilatation operator and remainder function.

\subsection{Dilatation operator}

The anomalous dilatation operator
is defined via the renormalisation factor $\cZ$ according to 
\begin{equation}\label{eq:dila}
\delta \dila=-\mu\frac{\dd}{\dd \mu}\log\cZ=2\peps g^2 \frac{\partial}{\partial g^2} \log{\cZ}\eqndot
\end{equation}
Expanding in powers of the coupling constant, we have
\begin{align}
\label{eq: expansion dilatation operator}
\delta \dila &=\sum_{\ell=1}^\infty g^{2\ell} \dila^{(\ell)}\eqncom
&
\cZ&=\sum_{\ell=0}^\infty g^{2\ell} \cZ^{(\ell)}\eqncom
&
\cZ^{(0)}&=\idm\eqndot
\end{align}

At one-loop order, the renormalised interaction (indicated by the underscore),
\begin{equation}
	\label{eq:renone-loopint}
	\Interactionr^{(1)}=\Interaction^{(1)}+\zone\eqncom
\end{equation}
is UV finite by definition of the renormalisation matrix $\cZ^{(1)}$. 
As $\Interaction^{(1)}$, we can expand $\zone$ in terms of densities as
\begin{equation}
\label{eq: zone in terms of densities}
 \cZ^{(1)}=\sum_{i=1}^L\zone[i\,i+1]\eqncom
\end{equation}
where we expand $\zone[i\,i+1]$ in terms of its matrix elements in complete analogy to \eqref{eq:one-loop-I-operator-form}:
\begin{equation}
\label{eq: Z one density in terms of matrix elements}
\zone[i\,i+1] = \sum_{n_i,m_i = 0}^{\infty} \zoneif{n_1, n_2}{m_1, m_2}  \frac{(p_i^+)^{m_1}}{m_1!} \frac{(p_{i+1}^+)^{m_2}}{m_2!} \bigg( {\partial \over \partial p^+_i} \bigg)_{p_i^+=0}^{n_1} \bigg( {\partial \over \partial p^+_{i+1}} \bigg)_{p_{i+1}^+=0}^{n_2}
\pnt
\end{equation}
By definition, $\zone[i\,i+1]$ has to cancel the UV divergence of $\interaction_{i\,i+1}$. In terms of matrix elements, this reads  
\begin{equation}
\label{eq: zone}
(\zone[i])_{n_1, n_2}^{m_1, m_2}=-(I_i^{(1)})_{n_1, n_2}^{m_1, m_2}\Big|_{\text{UV}}=-\Big[(I_i^{(1)})_{n_1, n_2}^{m_1, m_2}-(I_i^{(1)})_{n_1, n_2}^{m_1, m_2}\Big|_{\text{IR}}\Big]\Big|_{\frac{1}{\peps}\text{-pole}}\eqncom
\end{equation}
where we have written the UV divergence as the total divergence minus the IR divergence. 
The former is then given as the remaining ${\frac{1}{\peps}\text{-pole}}$, which we extract as indicated by the vertical bar.
The IR divergent part is universal  and given by (see e.g.\ \cite{Wilhelm:2014qua})
\begin{equation}
\label{IoneIR}
 (I_i^{(1)})_{n_1, n_2}^{m_1, m_2}\Big|_{\text{IR}}
 =-s_{i\,i+1}\delta_{n_1}^{m_1}\delta_{n_2}^{m_2}\FDinline[triangle,twolabels,labelone=\scriptscriptstyle p_i,labeltwo=\scriptscriptstyle p_{i+1}]
  =-\frac{1}{\peps^2}\left(-\frac{s_{i\,i+1}}{\mu^2}\right)^{-\peps}\delta_{n_1}^{m_1}\delta_{n_2}^{m_2}+O(\peps^0)\eqncom
\end{equation}
where the 
scalar
triangle integral is given in \eqref{eq:triangle}.

From equation \eqref{eq:dila}, we can read off the one-loop dilatation operator as
\begin{equation}
\dila^{(1)}=2 \peps \cZ^{(1)} \eqncom \quad \dilaone[i\,i+1]=2 \peps \zone[i\,i+1]\eqncom \quad \dilaoneif{n_1,n_2}{m_1,m_2}=2\peps\zoneif{n_1,n_2}{m_1,m_2}\eqncom
\end{equation}
where we have expanded the one-loop dilatation operator in terms of densities and matrix elements in complete analogy to \eqref{eq: zone in terms of densities} and \eqref{eq: Z one density in terms of matrix elements}, respectively. 
We have checked in explicit cases that the result from the above expression agrees with the well-known expression for the one-loop dilatation operator \cite{Beisert:2003jj}:
\begin{equation}
\label{eq:dilaone}
(\dila^{(1)})_{n_1,n_2}^{m_1,m_2}=2\left(S_{n_1}^{(1)}\delta_{n_1}^{m_1}\delta_{n_2}^{m_2}-
\frac{\theta_{n_2m_2}}{n_2-m_2}\delta_{n_1+n_2}^{m_1+m_2}
+
\left\{\begin{smallmatrix}
n_1\leftrightarrow n_2
\\
 m_1\leftrightarrow m_2
 \end{smallmatrix}
 \right\}\right)\eqndot
\end{equation}
Here, we employ the harmonic numbers defined by
\begin{align}
S_n^{(\ell)}&=\sum_{k=1}^n \frac{1}{k^\ell}
\end{align}
as well as the Heaviside function 
\begin{equation}
\theta_{nm}=
\left\{\begin{array}{lr}
        1 & \text{for } n>m\eqncom\\
        0 & \text{for } n\leq m\eqnsem
        \end{array}\right. 
\end{equation}
see also \cite{Belitsky:2006av} for the notation used here.

\subsection{Finite terms}

We would like to better understand the structure of the complete one-loop form factor obtained above. After extracting the dilatation operator as the $1/\peps$-pole of the interaction symbol $\cI$, let us now study the finite part of the form factor in the $\peps$-expansion.
We define the non-trivial finite part as the $\mathcal{O}(\peps^0)$ contributions
in the $\peps$-expansion of \eqref{eq:one-loop-integral-result} 
after the UV and IR divergences are subtracted using \eqref{eq: zone} and \eqref{IoneIR}:
\newcommand{\finite}{\operatorname{Fin}}
\begin{equation}
\label{eq:rational1}
 \finite\left(\intoneif{n_1,n_2}{m_1,m_2}\right)=\intoneif{n_1,n_2}{m_1,m_2}-\intoneif{n_1,n_2}{m_1,m_2}\Big|_{\text{IR}}+\zoneif{n_1, n_2}{m_1, m_2} \eqndot
\end{equation}
This yields in the special case $n_2=0$%
\footnote{Note that we could have absorbed the logarithm term by adding $\zoneif{n_1, n_2}{m_1, m_2}\Big(-\frac{s_{i\,i+1}}{\mu^2}\Big)^{-\peps}$ instead of $\zoneif{n_1, n_2}{m_1, m_2}$.}
\begin{align}
\finite\left(\intoneif{n_1,0}{m_1,m_2}\right)
&=\delta_{n_1}^{m_1}\delta_{0}^{m_2}\left[ -\frac{1}{2}\left(S_{n_1}^{(1)}\right)^2-\frac{3}{2}S_{n_1}^{(2)} 
\right] 
+\theta_{m_20}\delta_{n_1+0}^{m_1+m_2} \frac{1}{m_2}\left[ \frac{1}{m_2}+S_{n_1}^{(1)}
\right]\nonumber
\\*
&\phaneq+\frac{1}{2}\dilaoneif{n_1,0}{m_1,m_2}\log\big(-\tfrac{s_{i\,i+1}}{\mu^2}\big)
\label{eq: rational terms result}
\pnt
\end{align}
The general case can be obtained from this special case via the following symmetry relation for $\interaction^{(1)}_{i\,i+1}$:
\begin{equation}
\label{eq: one-loop momentum conservation}
m_1\intoneif{n_1,n_2}{m_1-1,m_2}+m_2\intoneif{n_1,n_2}{m_1,m_2-1}=(n_1+1)\intoneif{n_1+1,n_2}{m_1,m_2}+(n_2+1)\intoneif{n_1,n_2+1}{m_1,m_2}\eqncom
\end{equation}
which is a direct consequence of momentum conservation and leads to an identical equation for the finite part.
As a result, we find 
\begin{align}
\label{eq: rational terms general}
&\finite\left(\intoneif{n_1,n_2}{m_1,m_2}\right)=
\\* \nonumber
&{m_1!m_2! \over n_1! n_2!}\sum_{j=0}^{n_2}\sum_{k=0}^{n_2-j}\frac{(-1)^j n_2!}{j!k!(n_2-j-k)!} 
{(n_1+j)! \over (m_1-k)!(m_2-n_2+l)!}  
\finite\left(\intoneif{n_1+j, 0}{m_1-k, m_2-n_2+l}\right) \eqncom
\end{align}
where $l=j+k$.

We can see that the finite terms in \eqref{eq: rational terms result} and \eqref{eq: rational terms general} have a simple structure similar to the one of the dilatation operator in \eqref{eq:dilaone}.
In particular, they have uniform maximal transcendentality two if the harmonic numbers and inverse powers of differences of magnon numbers are attributed a degree of transcendentality according to the rules $S^{(l)}\rightarrow l, {1\over m_j-n_j}\rightarrow 1$. As we will see in the next section, the two-loop results show a similar pattern.

\section{Two loops}
\label{sec:two-loop}

In this section, we continue the study of SL(2) form factors at two loops. We will see that the unitarity method allows us to obtain the full integrand. We solve for the explicit dilatation operator and the remainder functions up to spin three. The two-loop dilatation operator and remainders show interesting patterns and suggest a hidden uniform maximal transcendentality property.

\subsection{Unitarity construction}

We first compute the two-loop integrand via the unitarity method. At two-loop order, connected interactions involve at most three fields of the composite operator, which are adjacent at the planar level. The two-loop corrections can be expressed as
\begin{equation}
\label{eq: two-loop interaction}
 \Interaction^{(2)}=\sum_{i=1}^L \Big( \inttwo[i\,i+1\,i+2] +\inttwo[i\,i+1] + \frac{1}{2} \sum_{j=i+2}^{L+i-2} \intone[i\,i+1] \intone[j\,j+1] \Big) \,,
\end{equation}
where $\inttwo[i\,i+1]$ and $\inttwo[i\,i+1\,i+2]$ denote connected interactions of range two and three, respectively.
We expand the interaction densities in terms of matrix elements as
\begin{align}
\inttwo[i\,i+1] &= \sum_{n_j,m_j = 0}^{\infty} (\interaction^{(2)}_i)_{n_1, n_2}^{m_1, m_2}  \frac{(p_i^+)^{m_1}}{m_1!} \frac{(p_{i+1}^+)^{m_2}}{m_2!}  
\bigg(\frac{\partial }{\partial p^+_i}\bigg)^{n_1}_{p^+_i=0}
\bigg(\frac{\partial }{\partial p^+_{i+1}}\bigg)^{n_2}_{p^+_{i+1}=0}
\col\label{eq: int two range two in matrix elements}\\
\inttwo[i\,i+1\,i+2] &= \sum_{n_j,m_j = 0}^{\infty} (\interaction^{(2)}_i)_{n_1, n_2,n_3}^{m_1, m_2,m_3}  \frac{(p_i^+)^{m_1}}{m_1!} \frac{(p_{i+1}^+)^{m_2}}{m_2!} \frac{(p_{i+2}^+)^{m_3}}{m_3!} \nonumber\\
&\hphantom{{}={}\sum_{n_j,m_j = 0}^{\infty} (\interaction^{(2)}_i)_{n_1, n_2,n_3}^{m_1, m_2,m_3}}
\bigg(\frac{\partial }{\partial p^+_i}\bigg)^{n_1}_{p^+_i=0}
\bigg(\frac{\partial }{\partial p^+_{i+1}}\bigg)^{n_2}_{p^+_{i+1}=0}
\bigg(\frac{\partial }{\partial p^+_{i+2}}\bigg)^{n_3}_{p^+_{i+2}=0}
\eqndot
\label{eq: int two range three in matrix elements}
\end{align}

The one-loop correction $(\interaction^{(1)}_i)_{n_1, n_2}^{m_1, m_2}$ has been given in the previous section. We apply the same unitarity technique to compute $(\interaction^{(2)}_i)_{n_1, n_2}^{m_1, m_2}$ and $(\interaction^{(2)}_i)_{n_1, n_2,n_3}^{m_1, m_2,m_3}$; the corresponding non-trivial double and triple cuts are shown in Fig.~\ref{fig: two-loop cuts}.
In the following, we briefly outline the major steps and present the full integrand. More technical details on this kind of calculation may be found in \cite{Loebbert:2015ova}.

\def\middletension{0.8}
\begin{figure}[htbp]	
\centering
\begin{subfigure}[t]{0.49\textwidth}
 \centering
$
\settoheight{\eqoff}{$\times$}%
\setlength{\eqoff}{0.5\eqoff}%
\addtolength{\eqoff}{-12.0\unitlength}%
\raisebox{\eqoff}{%
\fmfframe(2,2)(2,2){%
\begin{fmfchar*}(60,25)
\fmfleft{vp3,vp,vpL,vq}
\fmfright{vp2,vp1}
\fmf{dbl_plain_arrow,tension=1.2}{vq,v1}
\fmf{plain_arrow,tension=0}{v1,vpL}
\fmf{plain_arrow,tension=1.2}{v1,vp3}
\fmf{plain_arrow,left=0.7,label=$\scriptstyle l_1$,l.d=8,tension=0.4}{v1,v2}
\fmf{plain_arrow,right=0.7,label=$\scriptstyle l_2$,l.d=8,tension=0.4}{v1,v2}
\fmf{phantom_smallcut,left=0.7,tension=0.4}{v1,v2}
\fmf{phantom_smallcut,right=0.7,tension=0.4}{v1,v2}
\fmf{plain_arrow,tension=1.2}{v2,vp1}
\fmf{plain_arrow,tension=1.2}{v2,vp2}
\fmfv{decor.shape=circle,decor.filled=30,decor.size=30,label=$\scriptstyle \ffco_{\cO,,L}$,label.dist=0}{v1}
\fmfv{decor.shape=circle,decor.filled=0,decor.size=25}{v2}
\fmffreeze
\fmfdraw
 \fmfcmd{pair vertq, vertpone, vertptwo, vertpthree, vertpL, vertone, verttwo; vertone = vloc(__v1); verttwo = vloc(__v2); vertq = vloc(__vq); vertpone = vloc(__vp1); vertptwo = vloc(__vp2); vertpthree = vloc(__vp3);vertpL = vloc(__vpL);}
\fmfiv{decor.shape=circle,decor.filled=50,decor.size=35,label=$\scriptstyle \ampco_{4}$,label.dist=0}{verttwo}
\fmfiv{decor.shape=circle,decor.filled=0,decor.size=25,label=$\scriptstyle \ampco_{4}$,label.dist=0}{verttwo}
 \fmfiv{label=$\scriptstyle q$}{vertq}
 \fmfiv{label=$\scriptstyle p_1$}{vertpone}
 \fmfiv{label=$\scriptstyle p_2$}{vertptwo}
 \fmfiv{label=$\scriptstyle p_3$}{vertpthree}
 \fmfiv{label=$\scriptstyle p_L$}{vertpL}
 \fmfiv{label=$\cdot$,l.d=20,l.a=-150}{vertone}
 \fmfiv{label=$\cdot$,l.d=20,l.a=-165}{vertone}
 \fmfiv{label=$\cdot$,l.d=20,l.a=-180}{vertone}
\end{fmfchar*}%
}}%
$
\caption{First double cut in the $s_{12}$-channel.}
\label{fig: one two-loop double cut}
\end{subfigure}
%
\begin{subfigure}[t]{0.49\textwidth}
 \centering
$
\settoheight{\eqoff}{$\times$}%
\setlength{\eqoff}{0.5\eqoff}%
\addtolength{\eqoff}{-12.0\unitlength}%
\raisebox{\eqoff}{%
\fmfframe(2,2)(2,2){%
\begin{fmfchar*}(60,25)
\fmfleft{vp3,vp,vpL,vq}
\fmfright{vp2,vp1}
\fmf{dbl_plain_arrow,tension=1.2}{vq,v1}
\fmf{plain_arrow,tension=0}{v1,vpL}
\fmf{plain_arrow,tension=1.2}{v1,vp3}
\fmf{plain_arrow,left=0.7,label=$\scriptstyle l_1$,l.d=8,tension=0.4}{v1,v2}
\fmf{plain_arrow,right=0.7,label=$\scriptstyle l_2$,l.d=8,tension=0.4}{v1,v2}
\fmf{phantom_smallcut,left=0.7,tension=0.4}{v1,v2}
\fmf{phantom_smallcut,right=0.7,tension=0.4}{v1,v2}
\fmf{plain_arrow,tension=1.2}{v2,vp1}
\fmf{plain_arrow,tension=1.2}{v2,vp2}
\fmfv{decor.shape=circle,decor.filled=0,decor.size=25}{v1}
\fmfv{decor.shape=circle,decor.filled=50,decor.size=30,label=$\scriptstyle \ampco_{4}$,label.dist=0}{v2}
\fmffreeze
\fmfdraw
 \fmfcmd{pair vertq, vertpone, vertptwo, vertpthree, vertpL, vertone, verttwo; vertone = vloc(__v1); verttwo = vloc(__v2); vertq = vloc(__vq); vertpone = vloc(__vp1); vertptwo = vloc(__vp2); vertpthree = vloc(__vp3);vertpL = vloc(__vpL);}
\fmfiv{decor.shape=circle,decor.filled=30,decor.size=35,label=$\scriptstyle \ffco_{\cO,,L}$,label.dist=0}{vertone}
\fmfiv{decor.shape=circle,decor.filled=0,decor.size=25,label=$\scriptstyle \ffco_{\cO,,L}$,label.dist=0}{vertone}
 \fmfiv{label=$\scriptstyle q$}{vertq}
 \fmfiv{label=$\scriptstyle p_1$}{vertpone}
 \fmfiv{label=$\scriptstyle p_2$}{vertptwo}
 \fmfiv{label=$\scriptstyle p_3$}{vertpthree}
 \fmfiv{label=$\scriptstyle p_L$}{vertpL}
 \fmfiv{label=$\cdot$,l.d=20,l.a=-150}{vertone}
 \fmfiv{label=$\cdot$,l.d=20,l.a=-165}{vertone}
 \fmfiv{label=$\cdot$,l.d=20,l.a=-180}{vertone}
\end{fmfchar*}%
}}%
$
\caption{Second double cut in the $s_{12}$-channel.}
\label{fig: other two-loop double cut}
\end{subfigure}
\\[0.5\baselineskip]
\begin{subfigure}[t]{0.49\textwidth}
 \centering
$
\settoheight{\eqoff}{$\times$}%
\setlength{\eqoff}{0.5\eqoff}%
\addtolength{\eqoff}{-12.0\unitlength}%
\raisebox{\eqoff}{%
\fmfframe(2,2)(2,2){%
\begin{fmfchar*}(60,25)
\fmfleft{vp4,vp,vpL,vq}
\fmfright{vp3,vp2,vp1}
\fmf{dbl_plain_arrow,tension=1.2}{vq,v1}
\fmf{plain_arrow,tension=0}{v1,vpL}
\fmf{plain_arrow,tension=1.2}{v1,vp4}
\fmf{plain_arrow,left=0.7,label=$\scriptstyle l_1$,l.d=8,tension=0.8}{v1,v2}
\fmf{plain_arrow,right=0.7,label=$\scriptstyle l_3$,l.d=8,tension=0.8}{v1,v2}
\fmf{phantom_smallcut,left=0.7,tension=0}{v1,v2}
\fmf{phantom_smallcut,right=0.7,tension=0}{v1,v2}
\fmf{plain_arrow,tension=0,tag=1}{v1,v2}
\fmf{phantom_smallcut,tension=0}{v1,v2}
\fmf{plain_arrow,tension=1.2}{v2,vp1}
\fmf{plain_arrow,tension=0}{v2,vp2}
\fmf{plain_arrow,tension=1.2}{v2,vp3}
\fmfv{decor.shape=circle,decor.filled=30,decor.size=30,label=$\scriptstyle \ffco_{\cO,,L}$,label.dist=0}{v1}
\fmfv{decor.shape=circle,decor.filled=50,decor.size=30,label=$\scriptstyle \ampco_{6}$,label.dist=0}{v2}
\fmffreeze
 \fmfcmd{pair vertq, vertpone, vertptwo, vertpthree, vertpfour, vertpL, vertone, verttwo; vertone = vloc(__v1); verttwo = vloc(__v2); vertq = vloc(__vq); vertpone = vloc(__vp1); vertptwo = vloc(__vp2); vertpthree = vloc(__vp3); vertpfour = vloc(__vp4);vertpL = vloc(__vpL);}
 \fmfiv{label=$\scriptstyle q$}{vertq}
 \fmfiv{label=$\scriptstyle p_1$}{vertpone}
 \fmfiv{label=$\scriptstyle p_2$}{vertptwo}
 \fmfiv{label=$\scriptstyle p_3$}{vertpthree}
 \fmfiv{label=$\scriptstyle p_4$}{vertpfour}
 \fmfiv{label=$\scriptstyle p_L$}{vertpL}
 \fmfiv{label=$\cdot$,l.d=20,l.a=-150}{vertone}
 \fmfiv{label=$\cdot$,l.d=20,l.a=-165}{vertone}
 \fmfiv{label=$\cdot$,l.d=20,l.a=-180}{vertone}
\fmfipath{p[]}
\fmfiset{p1}{vpath1(__v1,__v2)}
\fmfiv{label=$\scriptstyle l_2\,,$,l.d=5,l.a=-115}{point length(p1)/2 of p1}
\end{fmfchar*}%
}}%
$
\caption{Triple cut in the $s_{123}$-channel.}
\label{fig: two-loop p1+p2+p3 triple cut}
\end{subfigure}
%
\begin{subfigure}[t]{0.49\textwidth}
 \centering
$
\settoheight{\eqoff}{$\times$}%
\setlength{\eqoff}{0.5\eqoff}%
\addtolength{\eqoff}{-12.0\unitlength}%
\raisebox{\eqoff}{%
\fmfframe(2,2)(2,2){%
\begin{fmfchar*}(60,25)
\fmfleft{vp3,vp,vpL,vq}
\fmfright{vp2,vp1}
\fmf{dbl_plain_arrow,tension=1.2}{vq,v1}
\fmf{plain_arrow,tension=0}{v1,vpL}
\fmf{plain_arrow,tension=1.2}{v1,vp3}
\fmf{plain_arrow,left=0.7,label=$\scriptstyle l_1$,l.d=8,tension=0.4}{v1,v2}
\fmf{plain_arrow,right=0.7,label=$\scriptstyle l_3$,l.d=8,tension=0.4}{v1,v2}
\fmf{phantom_smallcut,left=0.7,tension=0.4}{v1,v2}
\fmf{phantom_smallcut,right=0.7,tension=0.4}{v1,v2}
\fmf{plain_arrow,tension=0,tag=1}{v1,v2}
\fmf{phantom_smallcut,tension=0}{v1,v2}
\fmf{plain_arrow,tension=1.2}{v2,vp1}
\fmf{plain_arrow,tension=1.2}{v2,vp2}
\fmfv{decor.shape=circle,decor.filled=30,decor.size=30,label=$\scriptstyle \ffco_{\cO,,L+1}$,label.dist=0}{v1}
\fmfv{decor.shape=circle,decor.filled=50,decor.size=30,label=$\scriptstyle \ampco_{5}$,label.dist=0}{v2}
\fmffreeze
 \fmfcmd{pair vertq, vertpone, vertptwo, vertpthree, vertpL, vertone, verttwo; vertone = vloc(__v1); verttwo = vloc(__v2); vertq = vloc(__vq); vertpone = vloc(__vp1); vertptwo = vloc(__vp2); vertpthree = vloc(__vp3);vertpL = vloc(__vpL);}
 \fmfiv{label=$\scriptstyle q$}{vertq}
 \fmfiv{label=$\scriptstyle p_1$}{vertpone}
 \fmfiv{label=$\scriptstyle p_2$}{vertptwo}
 \fmfiv{label=$\scriptstyle p_3$}{vertpthree}
 \fmfiv{label=$\scriptstyle p_L$}{vertpL}
 \fmfiv{label=$\cdot$,l.d=20,l.a=-150}{vertone}
 \fmfiv{label=$\cdot$,l.d=20,l.a=-165}{vertone}
 \fmfiv{label=$\cdot$,l.d=20,l.a=-180}{vertone}
\fmfipath{p[]}
\fmfiset{p1}{vpath1(__v1,__v2)}
\fmfiv{label=$\scriptstyle l_2\,,$,l.d=5,l.a=-115}{point length(p1)/2 of p1}
\end{fmfchar*}%
}}%
$
\caption{Triple cut in the $s_{12}$-channel.}
\label{fig: two-loop p1+p2 triple cut}
\end{subfigure}
\caption{Unitary cuts of the two-loop form factor $\ffco_{\cO,L}^{(2)}$.}
\label{fig: two-loop cuts}
\end{figure}

\paragraph{Double cuts.}
The simplest cuts are the two-particle cuts in the two-particle channels, with the one-loop amplitude or form factor on one side of the cut. They are shown in Fig.~\ref{fig: one two-loop double cut} and \ref{fig: other two-loop double cut}, respectively.
From them, we can fix the integrand contributions from the following topologies as
\begin{align}
(\interaction^{(2)}_{i,A})_{n_1,n_2} = &\FDinline[rainbow,twolabels,labelone=\scriptscriptstyle i,labeltwo=\scriptscriptstyle i+1,momentum]
s_{i\,i+1}^2 \frac{(l_1^{+})^{n_1}}{n_1!}\frac{(l_2^+)^{n_{2}}}{n_2!} \eqncom
\label{eq:integrand-planar-ladder}
\\
(\interaction^{(2)}_{i,B})_{n_1,n_2,n_3} =   & \FDinline[trianglebox,dermomenta,threelabels,labelone=\scriptscriptstyle i,labeltwo=\scriptscriptstyle i+1,labelthree=\scriptscriptstyle i+2]
s_{l_1\,l_2}s_{i+1\,i+2}\frac{(l_1^{+})^{n_1}}{n_1!}\frac{(l_2^+)^{n_{2}}}{n_2!}\frac{(l_3^{+})^{n_{3}}}{n_3!} \eqncom
\label{eq:integrand-tribox}
\\
(\interaction^{(2)}_{i,C})_{n_1,n_2,n_3} = & \FDinline[boxtriangle,dermomenta,threelabels,labelone=\scriptscriptstyle i,labeltwo=\scriptscriptstyle i+1,labelthree=\scriptscriptstyle i+2] 
s_{i\,i+1}s_{l_2\,l_3}
\frac{(l_1^{+})^{n_1}}{n_1!}\frac{(l_2^+)^{n_{2}}}{n_2!}\frac{(l_3^{+})^{n_{3}}}{n_3!} \eqncom
\label{eq:integrand-boxtri}
\end{align}
where $n_1, n_{2}, n_{3}$ are the numbers of derivatives occurring in the corresponding minimal tree-level form factor and thus in the integrand. 
Moreover, we have abbreviated $l_2=p_i+p_{i+1}-l_1$ in $\interaction^{(2)}_{i,A}$ and $l_2=p_i+p_{i+1}+p_{i+2}-l_1-l_3$ in $\interaction^{(2)}_{i,B}$ as well as $\interaction^{(2)}_{i,C}$.
Note that these contributions can be equally obtained via consecutive double cuts.

\paragraph{Triple cut in the three-particle channel.}

Next, we perform the triple cut in the three-particle channel as shown in Fig.~\ref{fig: two-loop p1+p2+p3 triple cut}. Both \eqref{eq:integrand-tribox} and \eqref{eq:integrand-boxtri} contribute to this cut. The new contribution determined by this cut is
\begin{align}
(\interaction^{(2)}_{i,D})_{n_1,n_2,n_3} = 
-\FDinline[dermomenta,doubletrianglethree,threelabels,labelone=\scriptscriptstyle i,labeltwo=\scriptscriptstyle i+1,labelthree=\scriptscriptstyle i+2]
s_{i\,i+1\,i+2}
\frac{(l_1^{+})^{n_1}}{n_1!}\frac{(l_2^+)^{n_{2}}}{n_2!}\frac{(l_3^{+})^{n_{3}}}{n_3!} \pnt
\label{eq:integrand-tritri}
 \end{align}

\paragraph{Triple cut in the two-particle channel.}

The most complicated cut is the triple cut in the two-particle channel as shown in Fig.~\ref{fig: two-loop p1+p2 triple cut}. All previous integrands contribute to this cut. The new integrand contributions are 
\begin{align}
&
(\interaction^{(2)}_{i,E})_{n_1,n_2,n_3}  =  
\Biggl(\FDinline[trianglesubbubbletop,dermomenta,twolabels,labelone=\scriptscriptstyle i+1,labeltwo=\scriptscriptstyle i+2,labelthree=\scriptscriptstyle i+3](p_{i+1}^++p_{i+2}^+)
        -
   \FDinline[doubletriangletwo,dermomenta,threelabels,labelone=\scriptscriptstyle i+1,labeltwo={},labelthree=\scriptscriptstyle i+2]p_{i+1}^{+}s_{l_2\,l_3}
     \nonumber\\*
     & \,\,\,
-\FDinline[trianglefishtop,dermomenta,twolabels,labelone=\scriptscriptstyle i+1,labeltwo=\scriptscriptstyle i+2,labelthree=\scriptscriptstyle i+3] s_{i+1\,i+2}(l_1^++l_{2}^+)
   \Biggr) 
   \Biggl[{}\sum _{m=1}^{n_{1}} \frac{1}{m}
 \frac{(p_{i}^{+})^{n_{1}-m}}{(n_1-m)!}\frac{(l_1^{+})^{m-1}}{(m-1)!} \frac{(l_2^+)^{n_{2}}}{n_2!} \frac{(l_3^{+})^{n_{3}}}{n_3!}
 \\*
   &\,\,\,-\sum _{m=1}^{n_{2}} 
   \frac{1}{m} \frac{(p_{i}^{+})^{n_{1}}}{n_1!} \frac{(l_1^{+})^{m-1}}{(m-1)!} \frac{(l_2^+)^{n_{2}-m}}{(n_2-m)!} \frac{(l_3^{+})^{n_{3}}}{n_3!}
   -\sum _{m=1}^{n_{2}} 
   \frac{1}{m} \frac{(p_{i}^{+})^{n_{1}}}{n_1!}  \frac{(l_1^+)^{n_{2}-m}}{(n_2-m)!}\frac{(l_2^{+})^{m-1}}{(m-1)!} \frac{(l_3^{+})^{n_{3}}}{n_3!}\nonumber\\*
   &\,\,\,
   +\sum _{m=1}^{n_{3}} 
     \frac{1}{m}  \frac{(p_{i}^{+})^{n_{1}}}{n_1!}\frac{(l_1^{+})^{n_{2}}}{n_2!} \frac{(l_2^+)^{m-1}}{(m-1)!} \frac{(l_3^{+})^{n_{3}-m}}{(n_3-m)!}
     \Biggr] \eqncom \nonumber
\end{align}
and
\begin{align}
&
(\interaction^{(2)}_{i,F})_{n_1,n_2,n_3}  =   
        \Biggl(
        \FDinline[trianglesubbubblebottom,dermomenta,twolabels,labelone=\scriptscriptstyle i,labeltwo=\scriptscriptstyle i+1,labelthree=\scriptscriptstyle i+2](p_{i}^{+}+p_{i+1}^{+})
        -\FDinline[doubletriangletwo,dermomenta,threelabels,labelone=\scriptscriptstyle i,labeltwo={},labelthree=\scriptscriptstyle i+1]p_{i+1}^{+}s_{l_1\,l_2}
  \nonumber\\*
     &\,\,\, -\FDinline[trianglefishbottom,dermomenta,twolabels,labelone=\scriptscriptstyle i,labeltwo=\scriptscriptstyle i+1,labelthree=\scriptscriptstyle i+2] s_{i\,i+1}(l_2^{+}+l_{3}^{+}) 
     \Biggr) 
   \Biggl[{} 
   \sum _{m=1}^{n_{1}} 
 \frac{1}{m} \frac{(l_1^{+})^{n_{1}-m}}{(n_1-m)!}\frac{(l_2^+)^{m-1}}{(m-1)!} \frac{(l_3^{+})^{n_{2}}}{n_2!}  \frac{(p_{i+2}^{+})^{n_{3}}}{n_3!}
 \\*
 &\,\,\,
   -\sum _{m=1}^{n_{2}} 
   \frac{1}{m} \frac{(l_1^{+})^{n_{1}}}{n_1!} \frac{(l_2^+)^{n_{2}-m}}{(n_2-m)!}\frac{(l_3^{+})^{m-1}}{(m-1)!}  \frac{(p_{i+2}^{+})^{n_{3}}}{n_3!} 
  -\sum _{m=1}^{n_{2}} 
   \frac{1}{m} \frac{(l_1^{+})^{n_{1}}}{n_1!} \frac{(l_2^{+})^{m-1}}{(m-1)!}\frac{(l_3^+)^{n_{2}-m}}{(n_2-m)!}  \frac{(p_{i+2}^{+})^{n_{3}}}{n_3!} 
 \nonumber\\*
 &\,\,\,
+\sum _{m=1}^{n_{3}} 
\frac{1}{m} \frac{(l_1^{+})^{n_{1}}}{n_1!} \frac{(l_2^+)^{n_{2}}}{n_2!} \frac{(l_3^{+})^{m-1}}{(m-1)!}  \frac{(p_{i+2}^{+})^{n_{3}-m}}{(n_3-m)!}
      \Biggr] \eqncom  \nonumber
\end{align}
as well as
\begin{align}
&
(\interaction^{(2)}_{i,G})_{n_1,n_2}  =      
 \FDinline[doubletriangletwo,dermomenta,threelabels,labelone=\scriptscriptstyle i,labeltwo={},labelthree=\scriptscriptstyle i+1]s_{i\,i+1}
      \Biggl[    (p_{i+1}^{+}-l_3^+) \sum _{m=1}^{n_{1}} \frac{1}{m} \frac{(l_1^{+})^{n_1-m}}{(n_1-m)!} \frac{(l_2^+)^{m-1}}{(m-1)!}\frac{(l_3^{+})^{n_{2}}}{n_2!} 
\nonumber\\* 
&\,\,\,
       +(p_{i}^{+}-l_1^+) \sum _{m=1}^{n_{2}} \frac{1}{m} \frac{(l_1^{+})^{n_1}}{n_1!} \frac{(l_2^+)^{m-1}}{(m-1)!}\frac{(l_3^{+})^{n_{2}-m}}{(n_2-m)!} +\frac{(l_1^{+})^{n_{1}}}{n_1!} \frac{(l_3^{+})^{n_{2}}}{n_2!}
    \Biggr] \eqndot \label{eq:integrand-tritri-range2}
\end{align}
Here, we have $l_2=p_{i+1}+p_{i+2}-l_1-l_3$ in $\interaction^{(2)}_{i,E}$ and $l_2=p_i+p_{i+1}-l_1-l_3$ in $\interaction^{(2)}_{i,F}$ as well as $\interaction^{(2)}_{i,G}$.

We see that the structures of the loop momenta in the integrand are directly inherited from the corresponding structures in the tree-level form factors \eqref{eq:tree-minimal}--\eqref{eq: non-minimal replacement rule 3}.

\paragraph{Result.}
Given the above building blocks, the interaction densities can be defined as
\begin{equation}
\label{eq:two-loop-density-together}
 \begin{aligned}
\interaction^{(2)}_{i\,i+1}  & = \interaction^{(2)}_{i,A} +  \interaction^{(2)}_{i,G}   \,, \\
\interaction^{(2)}_{i\,i+1\,i+2}  & =  \interaction^{(2)}_{i,B} + \interaction^{(2)}_{i,C} + \interaction^{(2)}_{i,D} + \interaction^{(2)}_{i,E} + \interaction^{(2)}_{i,F}  \,. \\
\end{aligned}
\end{equation}
They can be expanded in terms of matrix elements as
\begin{equation}
 \begin{aligned}
  (\interaction^{(2)}_i)_{n_{1}, n_{2}}& = \sum_{m_1, m_{2}} (\interaction^{(2)}_i)_{n_{1}, n_{2}}^{m_1, m_{2}} \frac{(p_i^+)^{m_1}}{m_1!} \frac{(p_{i+1}^+)^{m_{2}}}{m_2!}  \,,\\
(\interaction^{(2)}_i)_{n_{1}, n_{2},n_3}& = \sum_{m_1, m_{2}, m_{3}} (\interaction^{(2)}_i)_{n_{1}, n_{2}, n_{3}}^{m_1, m_{2}, m_{3}} \frac{(p_i^+)^{m_1}}{m_1!} \frac{(p_{i+1}^+)^{m_{2}}}{m_2!} \frac{(p_{i+2}^+)^{m_{3}}}{m_3!} \,,
 \end{aligned}
\end{equation}
where $m_1+m_{2}=n_{1}+n_{2}$ and $m_1+m_{2}+m_{3}=n_{1}+n_{2}+n_{3}$, respectively.

Let us explain the special combination of building blocks in \eqref{eq:two-loop-density-together}. We should first note that there is an ambiguity in grouping and distributing the terms with integrals involving only two external momenta. 
In \eqref{eq:two-loop-density-together}, a choice is made such that the terms in $\interaction^{(2)}_{i,A}$ and
$\interaction^{(2)}_{i,G}$ are interpreted as contributions to the interaction density of range two acting on the sites $i$ and $i+1$, while the remaining terms are understood as interactions of range three acting on the sites $i$, $i+1$ and $i+2$. 
While this assignment is irrelevant when summing over all densities to obtain $\Interaction^{(2)}$, it ensures the finiteness of the remainder density obtained from the corresponding interaction densities. We will come back to this point in subsection \ref{sec:remainderstwoloop}, see the discussion after \eqref{eq: remainder density as graphs}.

As in the one-loop case, the integrands \eqref{eq:integrand-planar-ladder}--\eqref{eq:integrand-tritri-range2} contain uncontracted loop momenta.
Using PV reduction, they can be expressed in terms of Lorentz scalar integrands, which contain loop momenta in the numerator only in the form of irreducible scalar products.
The resulting integrals can be reduced to master integrals via IBP identities, using e.g.\ \texttt{LiteRed} \cite{Lee:2013mka}.
The required master integrals are all known analytically and can be found in \cite{Gehrmann:2000zt}. 
This procedure can be efficiently applied 
for small numbers of magnons; we have computed explicit results for all cases up to a total magnon number three, i.e.\ $\sum_in_i\leq3$. 
While in principle applicable to higher numbers of magnons, the computation time for the employed method significantly increases from four magnons on.

\subsection{Dilatation operator}

The two-loop renormalised form factor 
is given by
\begin{equation}
\label{eq: renormalised two-loop interaction}
\Interactionr^{(2)}= \Interaction^{(2)}+\Interaction^{(1)}\cZ^{(1)}+\cZ^{(2)} \eqncom
\end{equation}
where the two-loop renormalisation constant has to cancel the UV divergence and is given as
\begin{equation}
\label{eq: two-loop z}
\ztwo=\sum_{i=1}^L \Big( 
\ztwo_{i\,i+1\,i+2} + \frac{1}{2} \sum_{j=i+2}^{L+i-2} \zone[i\,i+1] \zone[j\,j+1] \Big) \eqndot
\end{equation}
The first term inside the summation above is due to the connected interactions, while the last term accounts for all disconnected products of one-loop interactions in \eqref{eq: two-loop interaction}.
Unlike in the one-loop case where the IR and UV divergences can be easily separated, at two-loop order the IR and UV divergences are entangled with each other. We can subtract the (universal) IR divergences via the BDS ansatz  \cite{Anastasiou:2003kj,Bern:2005iz}, as explained in detail in the next subsection. The two-loop renormalisation constant density $\ztwo_{i\,i+1\,i+2}$ can be obtained by imposing the BDS remainder density $\rem_{i\,i+1\,i+2}$ defined in \eqref{eq: remainder density as graphs} to be finite. 

In terms of the renormalisation matrix $\cZ$, the form of the two-loop dilatation operator follows from the definition \eqref{eq:dila}:
\begin{equation}\label{eq:two-loopD}
\dilatwo = 4 \peps \Big(\mathcal{Z}^{(2)} - {1\over2} (\mathcal{Z}^{(1)} )^2 \Big) \eqndot
\end{equation}
It can be expanded in terms of densities as 
\begin{equation}
\label{eq: two-loop dila in terms of densities}
 \dilatwo=\sum_{i=1}^L \dila_{i\,i+1\,i+2}^{(2)}\eqndot
\end{equation}
As usual, the densities are expressed via matrix elements 
in complete analogy to \eqref{eq: int two range three in matrix elements}.
Via the above relations, we may extract the matrix elements of the two-loop dilatation operator in analogy to the one-loop case. 
Note that the square in \eqref{eq:two-loopD} has to be evaluated as a square of matrices. As an example, consider the following matrix element of the dilatation operator density, which is given in terms of the (non-vanishing) matrix elements of $\mathcal{Z}^{(1)}$ and $\mathcal{Z}^{(2)}$ as:
\begin{align}
\label{eq:example-Z-product}
(\dila^{(2)}_i)_{0,1,2}^{2,1,0}=&4\peps
\bigg[
(\mathcal{Z}^{(2)}_i)_{0,1,2}^{2,1,0}
-\frac{1}{2}(\mathcal{Z}^{(1)}_i)_{0,\mathbf{3}}^{2,1}(\mathcal{Z}^{(1)}_{i+1})_{1,2}^{\mathbf{3},0}\bigg]
 \nonumber\\*
=&4\peps\bigg[
\bigg(\frac{1}{8\peps^2}-\frac{1}{16\peps}\bigg)-\frac{1}{2} \Big(-\frac{1}{2\peps}\Big)^2
\bigg]=-\frac{1}{4}\eqncom
\end{align}
where the boldface number indicates the magnon number at the single internal leg.
The matrix product will also be used later in the definition of the remainder function, see \eqref{eq: remainder density as graphs} for a pictorial representation. 
Instead of giving explicit results for all matrix elements up to three magnons, we may compare the obtained expressions to the previous literature, where the two-loop dilatation operator in the SL(2) sector can be found in closed form. 

\paragraph{Comparison with Feynman calculation.}

Notably, the full two-loop dilatation operator in the SL(2) sector of $\mathcal{N}=4$ SYM theory has been obtained by Belitsky, Korchemsky and Mueller in \cite{Belitsky:2006av}. The named authors provide an explicit expression in closed and relatively compact form. Our computation of the two-loop dilatation operator for states of up to three magnons shows perfect agreement with that operator if we evaluate both expressions on composite operators. 
For convenience, we display the respective expression in the following.%
\footnote{Note that the expression printed here differs from the expression given in \cite{Belitsky:2006av} by a conventional overall factor of $-2$ and by some conventional factorials of magnon numbers; we make these factorials explicit in the definitions of the matrix elements, which are defined in complete analogy to \eqref{eq: int two range two in matrix elements} and \eqref{eq: int two range three in matrix elements}. We also insert the correct factor of $2$ in front of $S_{n_1}^{(1)}S_{n_1}^{(2)}$ in \eqref{eq:Dragentwo1} compared to the expression given in the appendix of \cite{Belitsky:2006av}. We thank Grisha Korchemsky for communication and sharing related computer algebra files with us. }

In contrast to our expansion \eqref{eq: two-loop dila in terms of densities}, the authors of \cite{Belitsky:2006av} write the two-loop dilatation operator in terms of a range-two density and a range-three density:%
\footnote{Note that one may include the range-two density $\tilde{\dila}_{i\,i+1}^{(2)}$ into the definition of the range-three density $\tilde{\dila}_{i\,i+1\,i+2}^{(2)}$ by formally adding identity legs to $\tilde{\dila}_{i\,i+1}^{(2)}$.}
\begin{equation}
 \dilatwo=\sum_{i=1}^L\Big ( \tilde\dila_{i\,i+1}^{(2)}+\tilde\dila_{i\,i+1\,i+2}^{(2)}\Big)\eqndot
\end{equation}
The range-two density reads
\begin{align}\label{eq:Dragentwo1}
(\tilde\dila_i^{(2)})&_{n_1,n_2}^{m_1,m_2}
=
-2\Big[ 2S_{n_1}^{(1)}S_{n_1}^{(2)}+S_{n_1}^{(3)}\Big]\delta_{n_1}^{m_1}\delta_{n_2}^{m_2}
+
\frac{\theta_{n_2m_2}}{n_2-m_2}\delta_{n_1+n_2}^{m_1+m_2}
\Big[
S_{n_1}^{(2)}+3S_{m_1}^{(2)}
\nonumber\\*
&\qquad\qquad
+\Big(S_{m_1}^{(1)}-S_{n_1}^{(1)}\Big)\Big(S_{n_1}^{(1)}+3 S_{m_1}^{(1)}-4 S_{n_2-m_2-1}^{(1)}\Big)
\Big]
+
\left\{\begin{smallmatrix}
n_1\leftrightarrow n_2
\\
 m_1\leftrightarrow m_2
 \end{smallmatrix}
 \right\}\eqncom
\end{align}
and the range-three density reads (for $n_2=0$) 
\begin{align}
\label{eq: dilatation 2-loop range 3 0}
(\tilde\dila_i^{(2)})_{n_1,0,n_3}^{m_1,m_2,m_3}
&=
-2
\frac{\delta_{n_1+0+n_3}^{m_1+m_2+m_3}\theta_{n_3m_3}}{(n_1-m_1)(n_3-m_3)}\bigg[ \theta_{m_1n_1}\Big(\sfrac{1}{n_1-m_1}+S_{m_1}^{(1)}-S_{n_1}^{(1)}-S_{m_1-n_1}^{(1)}\Big)
\nonumber\\*
&\qquad\qquad+\theta_{n_1m_1}\Big(S_{n_1-m_1}^{(1)}+S_{n_3-m_3}^{(1)}-S_{m_2}^{(1)}\Big)\bigg]
+
\left\{\begin{smallmatrix}
n_1\leftrightarrow n_3
\\
 m_1\leftrightarrow m_3
 \end{smallmatrix}
 \right\},
\end{align}
as well as (for $n_2\neq 0$)
\begin{align}
\label{eq:dilatation-2loop-general}
(\tilde\dila_i^{(2)})_{n_1,n_2,n_3}^{m_1,m_2,m_3}=
&
\frac{m_1!m_2!m_3!}{n_1!n_2!n_3!}
\sum_{j_1=0}^{n_2}\sum_{j_3=0}^{n_2-j_1}\sum_{k_1=0}^{n_2-j}\sum_{k_3=0}^{n_2-j-k_1}\frac{(-1)^j 
n_2!
}{j_1!j_3!k_1!k_3!(n_2-\ell)!} 
\nonumber\\*
&{(n_1+j_1)! (n_3+j_3)! \over (m_1-k_1)! (m_2-n_2+\ell)! (m_3-k_3)!}(\tilde\dila_i^{(2)})_{n_1+j_1,0,n_3+j_3}^{m_1-k_1,m_2-n_2+\ell,m_3-k_3},
\end{align}
where
$j=j_1+j_3$ and $\ell=j+k_1+k_3$.
Note that the match
with the two-loop dilatation operator of \cite{Belitsky:2006av} is only at the level of the composite operators, i.e.\ after we sum over all matrix elements of the dilatation operator density that occur for a given (cyclically symmetric) single-trace operator. The matrix elements of the dilatation operator density do not directly match with the expressions of  \cite{Belitsky:2006av}, which is of course also not required. To make this clear, we have marked the matrix elements of \cite{Belitsky:2006av} with tildes.

We notice that the matrix elements \eqref{eq:Dragentwo1} and \eqref{eq: dilatation 2-loop range 3 0} of the two-loop dilatation operator are of uniform transcendentality three if we formally assign the transcendentality $k$ to the harmonic number $S^{(k)}$ and the transcendentality one to the fraction $\frac{1}{m_j-n_j}$ for $j=1,2,3$.
With the same assignment, the matrix elements \eqref{eq:dilaone} of the one-loop dilatation operator are formally of uniform transcendentality one. Let us refer to this property as hidden uniform transcendentality.
Note that the assignment of transcendentality to harmonic numbers is standard in the context of anomalous dimensions, see e.g.\ \cite{Kotikov:2002ab}. 
An assignment of transcendentality to similar fractions also occurred in other contexts in the literature \cite{Fleischer:1997bw,Fleischer:1998nb,Bianchi:2013sta}.

\paragraph{Comparison with algebraic construction.}
In \cite{Zwiebel:2008gr}, Zwiebel introduced an algebraic construction that allows to recursively compute the two-loop dilatation operator within the SL(2) sector. We have implemented this loop-recursion and we compared the spectrum of the resulting operator to the above findings for the $M=1,2,3$ excitation sectors. In all cases, we found perfect agreement. 
The operatorial form of the two-loop dilatation operator obtained by the above recursion is rather lengthy and we thus refrain from displaying it here. We note, however, that the recursion in \cite{Zwiebel:2008gr} actually applies to the larger $\mathrm{PSU}(1,1|2)$ sector of $\mathcal{N}=4$ SYM theory, which makes it interesting for future extensions of our work. Another advantage of this algebraic method is that it allows to construct a dilatation operator that is manifestly hermitian.%
\footnote{Note that in perturbation theory loop corrections to the dilatation operator do not have to be hermitian, see e.g.\ \cite{Braun:2001qx,Pomoni:2011jj}.}
\subsection{Remainders}
\label{sec:remainderstwoloop}

The (two-loop) BDS remainder \cite{Anastasiou:2003kj,Bern:2005iz} in the refined case of renormalised form factors \cite{Loebbert:2015ova} takes the form 
\begin{equation}
\label{eq: remainder}
 \Rem=\Interactionr^{(2)}(\peps) - \frac{1}{2}\left(\Interactionr^{(1)}(\peps)\right)^2 -  f^{(2)}(\peps)\Interactionr^{(1)}(2\peps)  
 + \cO (\peps ) \eqncom
 \end{equation}
where 
\begin{equation}
 f^{(2)} (\peps)= - 2 \zeta_2  - 2 \zeta_3 \peps - 2 \zeta_4 \peps^2 \eqndot 
\end{equation}
Similar to the example discussed in \eqref{eq:example-Z-product}, the product of the one-loop density with itself has to be understood as a matrix product. 
For the two-loop remainder
\begin{equation}
\label{eq: sum of remainder densities}
\Rem = 
\sum_{i=1}^L \rem_{i\,i+1\,i+2} \eqncom
\end{equation}
this is captured by the following pictorial 
expression
\begin{align}
\label{eq: remainder density as graphs}
 &\rem_{i\,i+1\,i+2} = \\*&\nonumber
\phaneq\frac{1}{2} 
 \begin{aligned}
 \begin{tikzpicture}
  \drawvlineblob{1}{1}
  \drawvlineblob{2}{1}
  \drawvlineblob{3}{1}
  \drawtwoblob{1}{1}{$\inttwoi$}
 \end{tikzpicture}
 \end{aligned}
 +
 \begin{aligned}
 \begin{tikzpicture}
  \drawvlineblob{1}{1}
  \drawvlineblob{2}{1}
  \drawvlineblob{3}{1}
  \drawthreeblob{1}{1}{$\inttwoi$}
 \end{tikzpicture}
 \end{aligned}
 +
 \frac{1}{2} 
 \begin{aligned}
 \begin{tikzpicture}
  \drawvlineblob{1}{1}
  \drawvlineblob{2}{1}
  \drawvlineblob{3}{1}
  \drawtwoblob{2}{1}{$\inttwoip$}
 \end{tikzpicture}
 \end{aligned}
 +  
 \begin{aligned}
 \begin{tikzpicture}
  \drawvlineblob{1}{1}
  \drawvlineblob{2}{1}
  \drawvlineblob{3}{1}
  \drawthreeblob{1}{1}{$\ztwoi$}
 \end{tikzpicture}
 \end{aligned}
 +
 \frac{1}{2} 
 \begin{aligned}
 \begin{tikzpicture}
  \drawvlineblob{1}{2}
  \drawvlineblob{2}{2}
  \drawvlineblob{3}{2}
  \drawtwoblob{1}{1}{$\zonei$}
  \drawtwoblob{1}{2}{$\intonei$}
 \end{tikzpicture}
 \end{aligned}
 +
 \begin{aligned}
 \begin{tikzpicture}
  \drawvlineblob{1}{2}
  \drawvlineblob{2}{2}
  \drawvlineblob{3}{2}
  \drawtwoblob{1}{1}{$\zonei$}
  \drawtwoblob{2}{2}{$\intoneip$}
 \end{tikzpicture}
 \end{aligned}
 +
 \begin{aligned}
 \begin{tikzpicture}
  \drawvlineblob{1}{2}
  \drawvlineblob{2}{2}
  \drawvlineblob{3}{2}
  \drawtwoblob{1}{2}{$\intonei$}
  \drawtwoblob{2}{1}{$\zoneip$}
 \end{tikzpicture}
 \end{aligned}
 + 
 \frac{1}{2} 
 \begin{aligned}
 \begin{tikzpicture}
  \drawvlineblob{1}{2}
  \drawvlineblob{2}{2}
  \drawvlineblob{3}{2}
  \drawtwoblob{2}{1}{$\zoneip$}
  \drawtwoblob{2}{2}{$\intoneip$}
 \end{tikzpicture}
 \end{aligned}
 \\*
 &\phaneq-\frac{1}{2}
 \left(
  \frac{1}{2} 
 \begin{aligned}
 \begin{tikzpicture}
  \drawvlineblob{1}{2}
  \drawvlineblob{2}{2}
  \drawvlineblob{3}{2}
  \drawtwoblob{1}{1}{$\intoneir$}
  \drawtwoblob{1}{2}{$\intoneir$}
 \end{tikzpicture}
 \end{aligned}
 +
 \begin{aligned}
 \begin{tikzpicture}
  \drawvlineblob{1}{2}
  \drawvlineblob{2}{2}
  \drawvlineblob{3}{2}
  \drawtwoblob{1}{1}{$\intoneir$}
  \drawtwoblob{2}{2}{$\intoneipr$}
 \end{tikzpicture}
 \end{aligned}
 +
 \begin{aligned}
 \begin{tikzpicture}
  \drawvlineblob{1}{2}
  \drawvlineblob{2}{2}
  \drawvlineblob{3}{2}
  \drawtwoblob{1}{2}{$\intoneir$}
  \drawtwoblob{2}{1}{$\intoneipr$}
 \end{tikzpicture}
 \end{aligned}
 + 
 \frac{1}{2} 
 \begin{aligned}
 \begin{tikzpicture}
  \drawvlineblob{1}{2}
  \drawvlineblob{2}{2}
  \drawvlineblob{3}{2}
  \drawtwoblob{2}{1}{$\intoneipr$}
  \drawtwoblob{2}{2}{$\intoneipr$}
 \end{tikzpicture}
 \end{aligned}
 \right)
 - f^{(2)}
 \left(
 \frac{1}{2} 
 \begin{aligned}
 \begin{tikzpicture}
  \drawvlineblob{1}{1}
  \drawvlineblob{2}{1}
  \drawvlineblob{3}{1}
  \drawtwoblob{1}{1}{$\intoneir$}
 \end{tikzpicture}
 \end{aligned}
 +
 \frac{1}{2} 
 \begin{aligned}
 \begin{tikzpicture}
  \drawvlineblob{1}{1}
  \drawvlineblob{2}{1}
  \drawvlineblob{3}{1}
  \drawtwoblob{2}{1}{$\intoneipr$}
 \end{tikzpicture}
 \end{aligned}
 \right)_{\peps\rightarrow2\peps}
 \eqndot
 \nonumber
\end{align}
The remainder density $\rem_{i\,i+1\,i+2}$ is expressed in terms of its matrix elements $\remif{n_1n_2n_3}{m_1m_2m_3}$ in complete analogy to \eqref{eq: int two range three in matrix elements}.
We determine $\ztwo_{i\,i+1\,i+2}$ from the requirement that the remainder density $\rem_{i\,i+1\,i+2}$ is finite.
This is possible only if momentum-dependent $\frac{1}{\peps}$-poles are absent in $\rem_{i\,i+1\,i+2}$ before adding $\ztwo_{i\,i+1\,i+2}$, for which 
it is crucial to have the correct assignment \eqref{eq:two-loop-density-together} of range and to average the range-two densities as in \eqref{eq: remainder density as graphs}. 

We have obtained explicit expressions of the remainder function \eqref{eq: remainder density as graphs} for all configurations up to magnon number three, i.e.\ $\remif{n_1,n_2,n_3}{m_1,m_2,m_3}$ with $n_1+n_2+n_3=m_1+m_2+m_3\leq3$.
The densities can be written as linear combinations of 14 functions, which are partially related by relabelling the kinematical variables. The latter also enter in the form of
\begin{equation}
\label{eq: uvw}
u_i = \frac{s_{i\,i+1}}{s_{i\,i+1 \, i+2}}\eqncom \quad
v_i = \frac{s_{i+1\,i+2}}{s_{i\,i+1 \, i+2}}\eqncom \quad
w_i = \frac{s_{i\,i+2}}{s_{i\,i+1 \, i+2}} \eqncom
\end{equation}
where 
\begin{equation}
s_{i\,i+1 \, i+2} = s_{i\,i+1}+s_{i+1\,i+2}+s_{i\,i+2} \eqndot
\end{equation}
Concretely, we can write 
\begin{equation}
\label{eq: remainder density in terms of coefficients}
 \remif{n_1,n_2,n_3}{m_1,m_2,m_3}=\cRem{n_1}{n_2}{n_3}{m_1}{m_2}{m_3}\cdot \vec{V}_i
 \equiv\sum_{k=1}^{14}[\cRem{n_1}{n_2}{n_3}{m_1}{m_2}{m_3}]_k[\vec{V}_i]_k\eqncom
\end{equation}
where $[\cRem{n_1}{n_2}{n_3}{m_1}{m_2}{m_3}]_k$ are rational functions of $u_i,v_i,w_i$ and
\begin{equation}
\label{eq:vectorV}
 \begin{aligned}
  \vec{V}_i=\biggl(&\remif{XXX}{XXX},
  \remif{YXX}{XYX}\Big|_{\textrm{deg-}3},
  \remif{XXY}{XYX}\Big|_{\textrm{deg-}3},
  -\frac{\pi ^2}{6}  \log \big(-\tfrac{s_{i+1,i+2}}{\mu^2}\big),
  -\frac{\pi ^2}{6}  \log \big(-\tfrac{s_{i,i+1}}{\mu^2}\big),\\
  &
  \log \big(\tfrac{u_i}{v_i}\big) \log \big(-\tfrac{s_{i\,i+1}}{\mu^2}\big)+\frac{\pi^2}{3},
  \log \big(\tfrac{v_i}{u_i}\big) \log \big(-\tfrac{s_{i+1\,i+2}}{\mu^2}\big)+\frac{\pi^2}{3},\\
  &\text{Li}_2(1-v_i)-\frac{1}{2} \log ^2(u_i)+\log (u_i)\log (v_i) -\frac{\pi ^2}{6},\\
  &\text{Li}_2(1-u_i)-\frac{1}{2} \log ^2(v_i)+\log (u_i)\log (v_i) -\frac{\pi ^2}{6},\\
 & -\frac{\pi ^2}{6},
  \log (u_i),
  \log (v_i),
  \log \big(-\tfrac{s_{i\,i+1\,i+2}}{\mu^2}\big),
  1
  \biggr)\eqndot
 \end{aligned}
\end{equation}
In the vector $\vec{V}_i$, the components are ordered according to 
their (decreasing) degree of transcendentality.
The first component of $\vec{V}_i$ has transcendentality four and is given by%
\footnote{Apart from classical polylogarithms, this expression contains the Goncharov polylogarithm $G\left(\left\{1-u_i,1-u_i,1,0\right\},v_i\right)$.}
 \begin{align}
 \remif{XXX}{XXX}\nonumber
 &=  -    \text{Li}_4(1-u_i)-\text{Li}_4(u_i)+\text{Li}_4\big(\tfrac{u_i - 1}{u_i}\big)
-  \log \big( \tfrac{1-u_i}{w_i }\big) 
\left[  \text{Li}_3\big(\tfrac{u_i - 1}{u_i}\big) - \text{Li}_3\big(1-u_i\big) \right] \\*\nonumber
 &\phaneq- \log \left(u_i\right) \left[\text{Li}_3\big(\tfrac{v_i}{1-u_i}\big)+\text{Li}_3\big(-\tfrac{w_i}{v_i}\big) + \text{Li}_3\big(\tfrac{v_i-1}{v_i}\big)
 -\frac{1}{3}  \log ^3\left(v_i\right) -\frac{1}{3} \log ^3\left(1-u_i\right)  \right]
 \\*\nonumber   
&\phaneq- \text{Li}_2\big(\tfrac{u_i-1}{u_i}\big) \text{Li}_2\big(\tfrac{v_i}{1-u_i}\big)+  \text{Li}_2\left(u_i\right) \left[
   \log \big(\tfrac{1-u_i}{ w_i }\big) \log \left(v_i \right) +\frac{1}{2} \log ^2\big( \tfrac{ 1-u_i }{ w_i }\big) \right] 
\\*\nonumber
&\phaneq+  \frac{1}{24} \log ^4\left(u_i\right)
-\frac{1}{8} \log ^2\left(u_i\right) \log ^2\left(v_i\right)  - \frac{1}{2} \log ^2\left(1-u_i\right) \log \left(u_i\right) \log \big( \tfrac{ w_i }{ v_i}\big)\\*\nonumber
&\phaneq- \frac{1}{2} \log \left(1-u_i\right) \log ^2\left(u_i\right) \log \left(v_i\right)
- \frac{1}{6} \log ^3\left(u_i\right) \log \left(w_i\right)  
  \\*\nonumber
  &\phaneq- \zeta_2 \Big[ \log \left(u_i\right) \log \big(\tfrac{1-v_i}{ v_i} \big)
+ \frac{1}{2}\log ^2\big( \tfrac{ 1-u_i }{ w_i }\big) - \frac{1}{2}\log ^2\left(u_i\right)   \Big]
\\*
&\phaneq + \zeta_3 \log (u_i) + \frac{\zeta_4}{ 2}  +G\left(\left\{1-u_i,1-u_i,1,0\right\},v_i\right) +  (u_i\, \leftrightarrow\, v_i) 
\pnt
 \end{align}
It is the density of the BPS remainder first obtained in \cite{Brandhuber:2014ica}. Moreover,
\begin{equation}
\label{eq: second matric element trans three}
\begin{aligned}
\remif{XXY}{XYX}\Big|_{\textrm{deg-}3}
&= \left[ \text{Li}_3\big(-\tfrac{u_i}{w_i}\big) - \log \left(u_i\right) \text{Li}_2\big(\tfrac{v_i}{1-u_i}\big)
+{1\over2} \log \left(1-u_i\right) \log \left(u_i\right) \log \big(\tfrac{w_i^2}{1-u_i} \big) \right.
\\*
&\phaneq \left. -\ {1\over2} \text{Li}_3\big(-\tfrac{u_i v_i}{w_i}\big) - {1\over2}\log \left(u_i\right) \log \left(v_i\right) \log \left(w_i\right) - \frac{1}{12} \log ^3\left(w_i\right) + (u_i\, \leftrightarrow\, v_i) \right]
\\*
&\phaneq -\ \text{Li}_3\left(1-v_i\right)+\text{Li}_3\left(u_i\right)-\frac{1}{2} \log ^2\left(v_i\right) \log \big(\tfrac{1-v_i}{u_i}\big)
+\frac{1}{6} \pi ^2 \log \big(\tfrac{v_i}{w_i}\big) 
\\*
& \phaneq
-\ \frac{1}{6} \pi ^2 \log \big(-\tfrac{s_{i\,i+1\,i+2}}{\mu^2}\big) 
\end{aligned}
\end{equation}
is the (up to relabelling) unique transcendentality-three contribution that occurs for remainders in the SU(2) sector built from the complex scalars $X$ and $Y$ \cite{Loebbert:2015ova}.
Here and below, ${}|_{\textrm{deg-}k}$ denotes the projection to the part of transcendentality degree $k$.
Note that at transcendentality two, besides the (up to relabelling) two functions encountered in the SU(2) sector 
\begin{equation}
 \begin{aligned}
\remif{XXY}{XYX}\Big|_{\textrm{deg-}2}&=\frac{1}{2}[\vec{V}_i]_6+[\vec{V}_i]_7+[\vec{V}_i]_8+[\vec{V}_i]_9-[\vec{V}_i]_{10}\eqncom\\
\remif{XXY}{XYX}\Big|_{\textrm{deg-}2}&=-\frac{1}{2}[\vec{V}_i]_7-[\vec{V}_i]_{9}\eqncom
 \end{aligned}
\end{equation}
only (up to relabelling) one further transcendentality-two function is required to express the transcendentality-two part of the remainder in the SL(2) sector.
A similar relation holds for the functions at transcendentality one.
Hence, compared to the SU(2) sector, an extended basis of 
functions with coefficients that contain ratios of Mandelstam variables
is required. This goes also beyond the SU(2$|$3) case, where in addition to 
the remainder densities in the SU(2) sector with rational coefficients only the constants $\pi^2$ and $\zeta_3$ were encountered in \cite{Brandhuber:2016fni}. The additional functions are still rather simple though.

One trivial symmetry of the remainder density is what is commonly called parity:
\begin{equation}
 \remif{n_1,n_2,n_3}{m_1,m_2,m_3}=\remif{n_3,n_2,n_1}{m_3,m_2,m_1}\Big|_{s_{i\,i+1}\leftrightarrow s_{i+1\,i+2},u_i\leftrightarrow v_i}\eqncom
\end{equation}
which is a simple relabelling.%
\footnote{Notice that the entries of $\vec{V}_i$ organise into singlets and doublets under parity.}
This symmetry is already visible at the level of the integrand of the two-loop interaction density.
Moreover, as a consequence of momentum conservation, the two-loop remainder obeys a relation analogous to
\eqref{eq: one-loop momentum conservation} at one loop. It reads
\begin{equation}
 \begin{aligned}
&m_1\remif{n_1,n_2,n_3}{m_1-1,m_2,m_3}+m_2\remif{n_1,n_2,n_3}{m_1,m_2-1,m_3}+m_3\remif{n_1,n_2,n_3}{m_1,m_2,m_3-1}\\
&=(n_1+1)\remif{n_1+1,n_2,n_3}{m_1,m_2,m_3}+(n_2+1)\remif{n_1,n_2+1,n_3}{m_1,m_2,m_3}+(n_3+1)\remif{n_1,n_2,n_3+1}{m_1,m_2,m_3}\pnt
\end{aligned}
\end{equation}
Using this identity, we can reconstruct $\remif{n_1,n_2,n_3}{m_1,m_2,m_3}$ from the knowledge of $ \remif{n_1,0,n_3}{m_1,m_2,m_3}$ in complete analogy to \eqref{eq:dilatation-2loop-general}:
\begin{align}
\label{eq:remainder-resum-identity}
 \remif{n_1,n_2,n_3}{m_1,m_2,m_3}
=
&
{m_1!m_2!m_3! \over n_1! n_2! n_3!}
\sum_{j_1=0}^{n_2}\sum_{j_3=0}^{n_2-j_1}\sum_{k_1=0}^{n_2-j}\sum_{k_3=0}^{n_2-j-k_1}\frac{(-1)^j 
n_2!}{j_1!j_3!k_1!k_3!(n_2-\ell)!}  
\nonumber\\*
&{(n_1+j_1)! (n_3+j_3)! \over (m_1-k_1)! (m_2-n_2+\ell)! (m_3-k_3)!}
\remif{n_1+j_1,0,n_3+j_3}{m_1-k_1,m_2-n_2+\ell,m_3-k_3} \eqncom
\end{align}
where
$j=j_1+j_3$ and $\ell=j+k_1+k_3$.
Thus, it is sufficient to look at $\remif{n_1,0,n_3}{m_1,m_2,m_3}$ with $n_1\geq n_3$, and $m_1>m_3$ if $n_1=n_3$. The corresponding coefficients $\cRem{n_1}{0}{n_3}{m_1}{m_2}{m_3}$ are given in appendix \ref{app: remainder densities}.

Inspecting the explicit expressions in the appendix, we observe several interesting patterns.
Firstly, the highest transcendental part is universal:
\begin{equation}
 [\cRem{n_1}{n_2}{n_3}{m_1}{m_2}{m_3}]_1=\delta_{n_1}^{m_1}\delta_{n_2}^{m_2}\delta_{n_3}^{m_3}\eqncom
\end{equation}
which is to say
\begin{equation}
 \remif{n_1,n_2,n_3}{m_1,m_2,m_3}\Big|_{\textrm{deg-}4}=\delta_{n_1}^{m_1}\delta_{n_2}^{m_2}\delta_{n_3}^{m_3}\remif{XXX}{XXX}\eqndot
\end{equation}
This property was already observed in \cite{Loebbert:2015ova} for the SU(2) sector and conjectured to extend to all operators.
Secondly, also the coefficients of the transcendentality-three functions $[\vec{V}_i]_{2}$ to  $[\vec{V}_i]_{5}$  follow some 
interesting patterns. 
We analyse the following expressions as examples below
\begin{align}
\label{eq: first pattern}
 \remif{m,0,0}{m,0,0}\Big|_{\textrm{deg-}3} &=- \sum_{k=1}^m\frac{1}{k}\Big(\frac{1-u_i}{w_i}\Big)^k \underbrace{\remif{YXX}{XYX}\Big|_{\textrm{deg-}3}}_{[\vec{V}_i]_2}-S^{(1)}_m\underbrace{(-1)\frac{\pi ^2}{6}  \log \big(-\tfrac{s_{i+1,i+2}}{\mu^2}\big)}_{[\vec{V}_i]_4} \eqncom\\
 \remif{m,0,0}{0,m,0}\Big|_{\textrm{deg-}3} &=\frac{1}{m}\underbrace{(-1)\frac{\pi ^2}{6}  \log \big(-\tfrac{s_{i+1,i+2}}{\mu^2}\big)}_{[\vec{V}_i]_4}\eqncom\\
 \remif{m,0,0}{0,0,m}\Big|_{\textrm{deg-}3} &=-\frac{(-1)^m}{m}\frac{u_i^{m}}{w_i^m}\underbrace{\remif{YXX}{XYX}\Big|_{\textrm{deg-}3}}_{[\vec{V}_i]_2}\eqncom\\
 \remif{m,0,0}{m-n,n,0}\Big|_{\textrm{deg-}3} &= \delta_{n,0}\remif{m,0,0}{m,0,0}\Big|_{\textrm{deg-}3}+\remif{n,0,0}{0,n,0}\Big|_{\textrm{deg-}3}\eqndot
\end{align}
Similar patterns are encountered also in the coefficients of the transcendentality-two functions; e.g.\ 
\begin{align}
 \remif{m,0,0}{0,m,0}\Big|_{\textrm{deg-}2} &=\frac{S^{(1)}_m}{2m}\underbrace{\left(\log \big(\tfrac{u_i}{v_i}\big) \log \big(-\tfrac{s_{i\,i+1}}{\mu^2}\big)+\frac{\pi^2}{3}\right)}_{[\vec{V}_i]_6}+\left(-\frac{S^{(1)}_m}{m}-\frac{1}{m^2}\right)\underbrace{(-1)\frac{\pi ^2}{6}}_{[\vec{V}_i]_{10}}\eqndot
\end{align}
We see that the coefficients of transcendental functions contain harmonic numbers and factors ${1\over m}$, where $\frac{1}{m}\equiv\frac{1}{m_j-n_j}$ for some $j$. These are precisely the building blocks occurring in the one- and two-loop dilatation operator \eqref{eq:dilaone}, \eqref{eq:Dragentwo1} and \eqref{eq: dilatation 2-loop range 3 0} and the one-loop finite terms \eqref{eq: rational terms result}. 
Moreover, we have generalisations of harmonic numbers which include ratios of Mandelstam variables: 
$\sum_{k=1}^m\frac{1}{k}\frac{(1-u_i)^{k}}{w_i^k}$. For $1-u_i=w_i$, the latter reduces to $S^{(1)}_m$.

The given matrix elements of the remainder function density have a formal transcendentality degree four if we attribute a degree to the above coefficients according to the following rules:
\begin{equation}\label{eq:replacerul}
S^{(l)}_m  \rightarrow l
\col\qquad
\sum_{k=1}^m\frac{1}{k}\frac{(1-u_i)^{k}}{w_i^k} \rightarrow 1
\col\qquad
\frac{1}{m_i-n_i} \rightarrow 1
\col\qquad
\frac{1}{m_j-n_j}\frac{u_i}{w_i} \rightarrow 1
\pnt
\end{equation}
This generalises the property of \emph{hidden maximal transcendentality} observed for the two-loop dilatation operator in the previous section to the case of functions of (ratios of) Mandelstam variables.
In the limit of large spin, we have 
\begin{equation}
\lim_{m\to\infty}S^{(l)}_m=\zeta_l
\col\qquad
\lim_{m\to\infty} \sum_{k=1}^m\frac{1}{k}\left(\frac{1-u_i}{w_i}\right)^k = -\log\big(1-\tfrac{1-u_i}{w_i}\big) \col
\end{equation}
i.e.\ the result is of uniform maximal transcendentality in the usual sense.

It would be very interesting to understand the full pattern of the remainders, in particular to obtain a closed expression in analogy to  \eqref{eq:Dragentwo1}--\eqref{eq:dilatation-2loop-general} for the dilatation operator.  In order to guess such an expression, we would need more data for higher numbers of magnons. We leave this point for future work.

\section{Conclusion and outlook}
\label{sec:conclusion}

In this paper, we have computed the minimal form factors of composite operators in the SL(2) sector of $\mathcal{N}=4$ SYM theory up to two-loop order. In particular, we have determined their finite remainders as well as the dilatation operator. The employed on-shell unitarity methods have once more proven to be an efficient tool in this context. 
Let us summarise the main results in some more detail:
\begin{itemize}
\item At one-loop order, we have calculated the full minimal form factor to all orders in the dimensional regularisation parameter. Extracting the one-loop dilatation operator from this data, we found perfect agreement with the existing literature.
Moreover, we have extracted the previously unknown finite terms.
\item At two loops, we have determined the two-loop integrand for states with an arbitrary number of magnons and the complete form factor remainder and dilatation operator for states with up to three magnons, i.e.\ with up to three derivative insertions into the composite operator. Also here, the obtained dilatation operator matches with the closed expression that was determined previously.
\item 
While the remainders contain functions of lower transcendentality, we have observed that their coefficients follow some interesting patterns. These support the conjecture of a hidden uniform maximal transcendentality property, cf.\ \secref{sec:intro}.
\end{itemize}

In order to confirm the above conjecture, the next logical step is to consider further examples of two-loop form factors.
In particular, an important goal is to extend the above results to all composite operators in the SL(2) sector, i.e.\ to composite operators with more than three magnons. 
This requires a more efficient approach to the PV and IBP reduction of the general integrals \eqref{eq:integrand-planar-ladder}--\eqref{eq:integrand-tritri-range2}.
Although it would be desirable to solve the integrals for arbitrary magnon numbers, this is not strictly necessary.
As the hidden uniform maximal transcendentality property conjectured above significantly constrains the dependence on the magnon numbers, we can fix the coefficients of an appropriate ansatz from explicit results with sufficiently high but finite magnon numbers. 
In order to better understand the transcendentality patterns, it may then be useful to consider the inverse Mellin transform of the remainder results.%
\footnote{For an example in the context of Wilson coefficients see e.g.\ \cite{Bianchi:2013sta}.}

It would also be interesting to consider larger subsectors closed under renormalisation, such as the PSU(1,1$|$2) sector. The two-loop dilatation operator in this subsector can be computed via the iterative construction given in \cite{Zwiebel:2008gr}, which should provide a useful guideline.
Finally, it would of course be very interesting to calculate the two-loop form factor remainder and the two-loop dilatation operator in the full theory, which is currently unknown.
We believe that the study of SL(2) form factors, with their new feature of non-compactness, 
provides an essential step in this direction.

While ${\cal N}=4$ SYM theory is interesting in its own right, methods developed in the context of this model have often lead to broad applications, such as in QCD or pure Yang--Mills theory.  
Apart from this, results from ${\cal N}=4$ SYM theory are also 
useful building blocks in QCD. 
Notably, based on the so-called maximal transcendentality principle 
\cite{Kotikov:2002ab}, the three-loop universal anomalous dimension was initially extracted \cite{Kotikov:2004er} as 
the leading transcendentality contribution of the three-loop non-singlet splitting functions of QCD \cite{Moch:2004pa}.\footnote{It was later 
tested by direct field-theory calculations \cite{Eden:2004ua,Sieg:2010tz}.}
Further evidence comes from the study of the three-point two-loop form factors of the stress-tensor supermultiplet in ${\cal N}=4$ SYM theory \cite{Brandhuber:2012vm}, where it was found that the two-loop remainder of  ${\cal N}=4$  SYM theory matches exactly the leading transcendental part of the two-loop Higgs-to-three-gluon amplitude in QCD \cite{Gehrmann:2011aa}. This generalises the maximal transcendentality principle from pure numbers to functions depending on the kinematics; see also \cite{Li:2016ctv}. 

Given these observations, it is plausible that the contribution to the QCD form factor remainder with the leading hidden transcendentality should coincide with the corresponding form factor remainder in ${\cal N}=4$ SYM theory. 
Again, the word `hidden' refers to the fact that besides harmonic numbers also factors such as $\frac{1}{m_j-n_j}$ 
should formally be attributed a degree of transcendentality. 
A possible starting point to check this conjecture would be a comparison between the two-loop dilatation operator of $\mathcal{N}=4$ SYM theory and QCD, see e.g.\ the recent work \cite{Braun:2016qlg}.
In \cite{Belitsky:2006av}, it was shown that the two-loop dilatation operator in the (scalar) SL(2) sector of $\mathcal{N}=2$ SYM theory is indeed given by the expression in $\mathcal{N}=4$ SYM theory plus terms of lower transcendentality, which at least points into this direction.
It would also be interesting to study the remainder functions of form factors in QCD with regard to this point.

\section*{Acknowledgments}

For helpful discussions and exchange, we would like to thank Lance Dixon, Burkhard Eden, Valentina Forini, Johannes Henn, Qing-Jun Jin, Thomas Klose, David Kosower, Anatoly Kotikov, Kasper Larsen, Dhritiman Nandan, Robert M.\ Schabinger, and in particular Grisha Korchemsky.
M.W.\ would like to thank the KITPC/ITP-CAS for kind hospitality during the final phase of this project.
The work of M.W.\ was supported   by  DFF-FNU  through grant number DFF-4002-00037. 
G.Y.\ was supported in part by the Chinese Academy of Sciences (CAS) Hundred-Talent Program and by the Key Research Program of Frontier Sciences of CAS (Grant No. QYZDB-SSW-SYS014), and by a DFG grant in the framework of the SFB 647 ``Raum-Zeit-Materie. Ana\-ly\-tische und Geometrische Strukturen''.
We also thank the Marie Curie network GATIS (gatis.desy.eu) of the European Union's Seventh Framework Programme FP7/2007-2013/ under REA Grant Agreement No 317089 for support.

\appendix

\section{Explicit one-loop integral}
\label{app: one-loop integral}

In this appendix, we evaluate the one-loop integral
\eqref{eq:one-loop-density-integrand} and show how to obtain 
the explicit result \eqref{eq:one-loop-integral-result}.

The integral \eqref{eq:one-loop-density-integrand}, shown here for $i=1$ for simplicity,
can be regarded as a certain component of the 
tensor integral
\begin{equation}\label{Imntp}
\begin{aligned}
(\interaction_1^{(1)})_{m, n,\mu_1\dots\mu_{m+n}}
&=-s_{12}(\e^{\gamma_{\text{E}}}\mu^2)^{2-\frac{D}{2}}\frac{1}{m!n!}\int\frac{\de^Dl_1}{i\pi^{\frac{D}{2}}}\frac{l_{1\,(\mu_1}\dots l_{1\,\mu_m}l_{2\,\mu_{m+1}}\dots l_{2\,\mu_{m+n})}}{l_1^2l_2^2(p_1-l_1)^2}\\
&=\sum_{k=0}^{m+n}(\interaction^{(1)}_1){}_{m,n}^{k,l}\frac{p_{1\,(\mu_1}\dots p_{1\,\mu_k}p_{2\,\mu_{k+1}}\dots p_{2\,\mu_{l+k})}}{k!l!}
\pnt
\end{aligned}
\end{equation}
Here, $l_2=p_1+p_2-l_1$, $s_{12}=(p_1+p_2)^2=2p_1\cdot p_2$, and
the parentheses around the Lorentz indices indicate traceless symmetrisation
w.r.t.\ the enclosed indices. 
The traceless symmetrised product built from a single momentum
was given in \cite{Kotikov:1995cw} as
\begin{equation}
\begin{aligned}\label{ltracelessprod}
l_{(\mu_1}\dots l_{\mu_n)}=\sum_{\sigma\in\mathds{S}_n}\sum_{j=0}^{[\frac{n}{2}]}
\frac{(-1)^j\Gamma(\tfrac{D}{2}+n-j-1)}{2^{2j}j!(n-2j)!\Gamma(\tfrac{D}{2}+n-1)}l^{2j}\eta_{\mu_{\sigma_1}\mu_{\sigma_2}}\dots \eta_{\mu_{\sigma_{2j-1}}\mu_{\sigma_{2j}}}l_{\mu_{\sigma_{2j+1}}}\dots l_{\mu_{\sigma_n}}\col
\end{aligned}
\end{equation}
where $\mathds{S}_{n}$ denotes the group of permutations of $n$ elements and $\eta_{\mu\nu}$ is the metric of $D$-dimensional Minkowski space in the mostly-minus convention. 
The above expression is identical to the ordinary product if the momentum 
$l$ is on-shell. Using the above result, an expression for the numerator of the 
integral in \eqref{Imntp} can be found:
\begin{equation}
\begin{aligned}
\label{l1l2tracelessprod}
&
l_{1\,(\mu_1}\dots l_{1\,\mu_m}l_{2\,\mu_{m+1}}\dots l_{2\,\mu_{m+n})}\\*
&
=\frac{1}{(m+n)!}\sum_{\sigma\in\mathds{S}_{m+n}}\sum_{j=0}^{\min(m,n)}
\frac{(-1)^jm!n!\Gamma(\tfrac{D}{2}+m+n-j-1)}{2^{j}j!(m-j)!(n-j)!\Gamma(\tfrac{D}{2}+m+n-1)}
(l_1\cdot l_2)^j\\
&\hphantom{{}={}\frac{1}{(m+n)!}\sum_{\sigma\in\mathds{Z}_{m+n}}}
\eta_{\mu_{\sigma_1}\mu_{\sigma_{m+1}}}\dots \eta_{\mu_{\sigma_{j}}\mu_{\sigma_{m+j}}}l_{1\,(\mu_{\sigma_{j+1}}}\dots l_{1\,\mu_{\sigma_{m}})}l_{2\,(\mu_{\sigma_{m+j+1}}}\dots l_{2\,\mu_{\sigma_{m+n}})}
\eqndot
\end{aligned}
\end{equation}
If the momenta $l_1$ and $l_2$ are on-shell, the traceless symmetric products
on the r.h.s.\ become ordinary products. 
It is easy to see that in this case the contraction of two of these tensor products 
with rank $m+n$ yields zero unless the two tensor products are related 
by exchanging $m$ and $n$.
This applies to the tensors in the decomposition 
in \eqref{Imntp}, and it allows us to project out the 
coefficients $(\interaction^{(1)}_1){}_{m,n}^{k,l}$ of \eqref{Imntp} by contractions with the tensors 
$p_2^{(\mu_1}\dots p_2^{\mu_k}p_1^{\mu_{k+1}}\dots p_1^{\mu_{k+l})}$, where
$k+l=m+n$:
\begin{equation}
\begin{aligned}
\label{Cmnklint}
(\interaction^{(1)}_1){}_{m,n}^{k,l}
&=-\frac{s_{12}^{1-m-n}}{P_{k\,l}}(\e^{\gamma_{\text{E}}}\mu^2)^{2-\frac{D}{2}}\frac{k!l!}{m!n!}\\
&\hphantom{{}={}}\int\frac{\de^Dl_1}{i\pi^{\frac{D}{2}}}\frac{l_{1\,(\mu_1}\dots l_{1\,\mu_m}l_{2\,\mu_{m+1}}\dots l_{2\,\mu_{m+n})}p_2^{\mu_1}\dots p_2^{\mu_k}p_1^{\mu_{k+1}}\dots p_1^{\mu_{k+l}}}{l_1^2l_2^2(p_1-l_1)^2}
\col
\end{aligned}
\end{equation}
where 
\begin{equation}
\begin{aligned}\label{Pmn}
P_{m,n}&=\frac{1}{s_{12}^{m+n}}p_{1\,(\mu_1}\dots p_{1\,\mu_m}p_{2\,\mu_{m+1}}\dots p_{2\,\mu_{m+n})}p_2^{\mu_1}\dots p_2^{\mu_m}p_1^{\mu_{m+1}}\dots p_1^{\mu_{m+n}}
=\frac{m!n!}{(m+n)!}Q_{m,n}\\
Q_{m,n}
&=\frac{1}{s_{12}^{m+n}}q_{(\mu_1}\dots q_{\mu_{m+n})}p_2^{\mu_1}\dots p_2^{\mu_m}p_1^{\mu_{m+1}}\dots p_1^{\mu_{m+n}}\\
&=-\frac{1}{2^{m+n}}
\frac{\Gamma(\tfrac{D}{2}+m-1)\Gamma(\tfrac{D}{2}+n-1)}{\Gamma(\tfrac{D}{2}+m+n-1)}\Gamma(2-\tfrac{D}{2})\frac{\sin\tfrac{\pi D}{2}}{\pi}
\pnt
\end{aligned}
\end{equation}
Here, $q=p_1+p_2$ with $q^2=s_{12}<0$ denotes an off-shell momentum which should not 
be confused with the total momentum used in the main text.
These latter expressions follow from contractions of \eqref{l1l2tracelessprod} and \eqref{ltracelessprod} in which 
$l_1$, $l_2$ are replaced by $p_1$, $p_2$ and $l$ is replaced by $q$, respectively.
The expression for $Q_{m,n}$, which is manifestly symmetric in $m$ and $n$ as
expected, was obtained by also making use of the reflection 
formula for the Gamma functions:
\begin{equation}\label{gammareflection}
\Gamma(z)\Gamma(1-z)=\frac{\pi}{\sin\pi z}\pnt
\end{equation}

The numerator in \eqref{Cmnklint} is given by contracting \eqref{l1l2tracelessprod}
with $p_2^{(\mu_1}\dots p_2^{\mu_k}p_1^{\mu_{k+1}}\dots p_1^{\mu_{k+l})}$, where $k+l=m+n$. This yields
\begin{equation}
\begin{aligned}
&l_{1\,(\mu_1}\dots l_{1\,\mu_m}l_{2\,\mu_{m+1}}\dots l_{2\,\mu_{m+n})}p_2^{\mu_1}\dots p_2^{\mu_k}p_1^{\mu_{k+1}}\dots p_1^{\mu_{k+l}}\\
&=\frac{m!n!}{(m+n)!}\sum_{j=0}^{\min(m,n,k,l)}
\frac{(-1)^jk!l!\Gamma(\tfrac{D}{2}+m+n-j-1)}{j!(k-j)!(l-j)!\Gamma(\tfrac{D}{2}+m+n-1)}
(l_1\cdot l_2)^j
(p_1\cdot p_2)^j\\
&\hphantom{{}={}}
\sum_{r=\max(k-n,0)}^{\min(m,k)-j}\binom{k-j}{r}\binom{l-j}{m-j-r}
l_{1\,(\nu_{1}}\dots l_{1\,\nu_{m-j})}p_2^{\nu_{1}}\dots p_2^{\nu_{r}}
p_1^{\nu_{r+1}}\dots p_1^{\nu_{m-j}}\\
&\hphantom{{}={}
\sum_{r=\max(k-n,0)}^{\min(m,k)-j}\binom{k-j}{r}\binom{l-j}{m-j-r}}
l_{2\,(\rho_{1}}\dots l_{2\,\rho_{n-j})}p_2^{\rho_{1}}\dots p_2^{\rho_{k-j-r}}
p_1^{\rho_{k-j-r+1}}\dots p_1^{\rho_{n-j}}
\pnt
\end{aligned}
\end{equation}
In the integral \eqref{Cmnklint}, all numerator terms that contain $l_1^2$
or $l_2^2$ lead to (massless) bubble integrals with a single on-shell momentum, which vanish. 
Hence, the traceless symmetric
products on the r.h.s.\ of the above expression can be regarded as ordinary products. Moreover, 
$l_1\cdot l_2$ can be replaced by $p_1\cdot p_2$. 
These changes do not alter the result of the integral, which can thus 
be expressed as
\begin{equation}\label{I1mnsums}
\begin{aligned}
&-s_{12}(\e^{\gamma_{\text{E}}}\mu^2)^{2-\frac{D}{2}}\int\frac{\de^Dl_1}{i\pi^{\frac{D}{2}}}\frac{l_{1\,(\mu_1}\dots l_{1\,\mu_m}l_{2\,\mu_{m+1}}\dots l_{2\,\mu_{m+n})}p_2^{\mu_1}\dots p_2^{\mu_k}p_1^{\mu_{k+1}}\dots p_1^{\mu_{k+l}}}{l_1^2l_2^2(p_1-l_1)^2}\\
&=\frac{m!n!}{(m+n)!}\sum_{j=0}^{\min(m,n,k,l)}
\frac{(-1)^jk!l!\Gamma(\tfrac{D}{2}+m+n-j-1)}{j!(k-j)!(l-j)!\Gamma(\tfrac{D}{2}+m+n-1)}
 \\
 &\hphantom{{}={}}
\sum_{r=\max(k-n,0)}^{\min(m,k)-j}\binom{k-j}{r}\binom{l-j}{m-j-r}
\sum_{s=0}^{k-j-r}\binom{k-j-r}{s}(-1)^s\\
&\phaneq\sum_{t=0}^{n-k+r}\binom{n-k+r}{t}(-1)^t
(p_1\cdot p_2)^{n+j-s-t}I_{r+s,m-j-r+t}
\col
\end{aligned}
\end{equation}
where we also have inserted $l_2=p_1+p_2-l_1$ and then expanded the obtained
binomial products.

The integral that occurs
in the above expression is given by
\begin{equation}
\begin{aligned}\label{Imn1}
I_{m,n}
&=-s_{12}(\e^{\gamma_{\text{E}}}\mu^2)^{2-\frac{D}{2}}\int\frac{\de^Dl_1}{i\pi^{\frac{D}{2}}}\frac{(l_1\cdot p_2)^m(l_1\cdot p_1)^n}{l_1^2l_2^2(p_1-l_1)^2}\\
&=-\Big(\frac{s_{12}}{2}\Big)^{m+n}
\frac{(-1)^{n}\Gamma(2-\tfrac{D}{2})}{\Gamma(3-n-\tfrac{D}{2})}\frac{\Gamma(D-2)\Gamma(\tfrac{D}{2}+m-1)}{\Gamma(\tfrac{D}{2}-1)\Gamma(D+m+n-3)}\FDinline[bubble,twolabels,labelone=\scriptscriptstyle p_1,labeltwo=\scriptscriptstyle p_{2}]
\col
\end{aligned}
\end{equation}
as explained below.
The diagram denotes the standard scalar 
bubble integral, i.e.\
\begin{equation}\label{B1}
\begin{aligned}
\FDinline[bubble,twolabels,labelone=\scriptscriptstyle p_1,labeltwo=\scriptscriptstyle p_{2}]&\equiv(\e^{\gamma_{\text{E}}}\mu^2)^{2-\frac{D}{2}}\int\frac{\de^Dl}{i\pi^{\frac{D}{2}}}\frac{1}{l^2(p_1+p_2-l)^2}\\
&=\e^{(2-\frac{D}{2})\gamma_{\text{E}}}\frac{\Gamma(\tfrac{D}{2}-1)^2\Gamma(2-\tfrac{D}{2})}{\Gamma(D-2)}\Big(\frac{\mu^2}{-s_{12}}\Big)^{2-\frac{D}{2}}\pnt
\end{aligned}
\end{equation}
In the case $n\ge1$, the result \eqref{Imn1} 
can straightforwardly be obtained by using
$l_1\cdot p_1=\frac{1}{2}(l_1^2-(l_1-p_1)^2)$
for one of the numerator factors and exploiting the fact that the 
cancellation of $l_1^2$ in the denominator yields 
a (massless) bubble integral with a single external on-shell momentum, 
which hence vanishes.
In the special case $n=0$ but $m\geq1$, one shifts the loop momentum as $l_1=\tilde{l}_1+p_1$
and expands the power of $(\tilde{l}_1+p_1)\cdot p_2$ in a binomial series, 
then using $\tilde{l}_1\cdot p_2=\frac{1}{2}(\tilde{l}_1^2-(\tilde{l}_1-p_2)^2)$ for one of the numerator 
factors of the resulting integrals. The resulting (massless)
bubble integrals
are either zero since they depend only on one external on-shell momentum or
they can be evaluated using (see e.g.\ \cite{Chetyrkin:1980pr})
\begin{equation}
\begin{aligned}\label{Gtracelessdef}
(\e^{\gamma_{\text{E}}}\mu^2)^{2-\frac{D}{2}}\int\frac{\de^Dl}{i\pi^{\frac{D}{2}}}\frac{l_{(\mu_1}\dots
  l_{\mu_n)}}{l^{2}(l-q)^{2}}
=q_{(\mu_1}\dots q_{\mu_n)}
\frac{\Gamma(\tfrac{D}{2}+n-1)}{\Gamma(D+n-2)}\frac{\Gamma(D-2)}{\Gamma(\tfrac{D}{2}-1)}\FDinline[bubble,twolabels,labelone=\scriptscriptstyle p_1,labeltwo=\scriptscriptstyle p_{2}]
\pnt
\end{aligned}
\end{equation}
Finally, also the scalar triangle diagram occurs, in particular for $n=m=0$ but also in the binomial expansions. It can be related to the bubble integral via
\begin{equation}\label{eq:triangle}
\FDinline[triangle,twolabels,labelone=\scriptscriptstyle p_1,labeltwo=\scriptscriptstyle p_{2}]=\frac{1}{s_{12}}\frac{2(D-3)}{4-D}\FDinline[bubble,twolabels,labelone=\scriptscriptstyle p_1,labeltwo=\scriptscriptstyle p_{2}]\pnt
\end{equation}
A resummation of the binomial series yields a result which is in accordance with \eqref{Imn1} for $n=0$.

After inserting \eqref{Imn1} into \eqref{I1mnsums}, the sums over $s$ and $t$ can be performed, and after also applying the reflection formula \eqref{gammareflection}, the expression reads
\begin{equation}
\begin{aligned}
&-s_{12}(\e^{\gamma_{\text{E}}}\mu^2)^{2-\frac{D}{2}}\int\frac{\de^Dl_1}{i\pi^{\frac{D}{2}}}\frac{l_{1\,(\mu_1}\dots l_{1\,\mu_m}l_{2\,\mu_{m+1}}\dots l_{2\,\mu_{m+n})}p_2^{\mu_1}\dots p_2^{\mu_k}p_1^{\mu_{k+1}}\dots p_1^{\mu_{k+l}}}{l_1^2l_2^2(p_1-l_1)^2}\\
&=-\Big(\frac{s_{12}}{2}\Big)^{m+n}\frac{m!n!}{(m+n)!}\frac{\Gamma(2-\tfrac{D}{2})\Gamma(D-2)}{\Gamma(\tfrac{D}{2}-1)}\frac{\sin\tfrac{\pi D}{2}}{\pi}\FDinline[bubble,twolabels,labelone=\scriptscriptstyle p_1,labeltwo=\scriptscriptstyle p_{2}]\\
&\phantom{{}={}}
\sum_{j=0}^{\min(m,n,k,l)}
\frac{(-1)^jk!l!\Gamma(\tfrac{D}{2}+m+n-j-1)}{j!(k-j)!(l-j)!\Gamma(\tfrac{D}{2}+m+n-1)}
\sum_{r=\max(0,l-n)}^{\min(l,m)-j}\binom{k-j}{m-j-r}\binom{l-j}{r}
\\
&\hphantom{{}={}\sum_{j=0}^{\min(m,n,k,l)}\sum_{r=\max(0,l-n)}^{\min(l,m)-j}}
\frac{\Gamma(\tfrac{D}{2}-m+k+2r-2)\Gamma(\tfrac{D}{2}+m+l-2j-2r-1)}{\Gamma(D+m+n-2j-3)}\eqndot
\end{aligned}
\end{equation}
Inserting this result and \eqref{Pmn} into \eqref{Cmnklint}, we finally obtain for the coefficients in \eqref{Imntp}:
\begin{equation}
\begin{aligned}
(\interaction_1^{(1)})_{m,n}^{k,l}
&=
\frac{k!l!}{\Gamma(\tfrac{D}{2}+k-1)\Gamma(\tfrac{D}{2}+l-1)}
\frac{\Gamma(D-2)}{\Gamma(\tfrac{D}{2}-1)}\FDinline[bubble,twolabels,labelone=\scriptscriptstyle p_1,labeltwo=\scriptscriptstyle p_{2}]\\
&\phantom{{}={}}
\sum_{j=0}^{\min(m,n,k,l)}
\frac{(-1)^j\Gamma(\tfrac{D}{2}+m+n-j-1)}{j!(k-j)!(l-j)!}
\sum_{r=\max(0,l-n)}^{\min(l,m)-j}\binom{k-j}{m-j-r}\binom{l-j}{r}
\\
&\hphantom{{}={}\sum_{j=0}^{\min(m,n,k,l)}\sum_{r=\max(0,l-n)}^{\min(l,m)-j}}
\frac{\Gamma(\tfrac{D}{2}-m+k+2r-2)\Gamma(\tfrac{D}{2}+m+l-2j-2r-1)}{\Gamma(D+m+n-2j-3)}\eqndot
\end{aligned}
\end{equation}

\section{Remainder densities}
\label{app: remainder densities}

In this appendix, we give the remainder densities up to spin three in the notation introduced in \eqref{eq: remainder density in terms of coefficients}.

\begin{dgroup*}
\begin{dmath*}
 \cRem{0}{0}{0}{0}{0}{0} = \left(1,0,0,0,0,0,0,0,0,0,0,0,0,0  \right)\end{dmath*}\begin{dmath*}
 \cRem{1}{0}{0}{0}{0}{1} = \left(0,\frac{u_i}{w_i},0,0,0,-\frac{(1-v_i)}{2 w_i},-\frac{u_i}{2 w_i},-\frac{u_i}{w_i}-1,-\frac{u_i}{w_i},-\frac{u_i}{w_i},1,-2,-1,\frac{7}{2}  \right)\end{dmath*}\begin{dmath*}
 \cRem{1}{0}{0}{0}{1}{0} = \left(0,0,0,1,0,\frac{1}{2},0,0,0,-2,0,2,2,-7  \right)\end{dmath*}\begin{dmath*}
 \cRem{1}{0}{0}{1}{0}{0} = \left(1,\frac{u_i}{w_i}-\frac{1}{w_i},0,-1,0,-\frac{(1-u_i)}{2 w_i},-\frac{(1-u_i)}{2 w_i},-\frac{v_i}{w_i},\frac{u_i}{w_i}-\frac{1}{w_i},\frac{(1-u_i)}{w_i}-\frac{2 v_i}{w_i},-1,0,-1,\frac{7}{2}  \right)\end{dmath*}\begin{dmath*}
 \cRem{1}{0}{1}{0}{2}{0} = \left(0,0,0,0,0,0,0,0,0,1,-\frac{1}{2},-\frac{1}{2},-1,\frac{15}{4}  \right)\end{dmath*}\begin{dmath*}
 \cRem{1}{0}{1}{1}{0}{1} = \left(1,\frac{u_i}{w_i}-\frac{1}{w_i},-\frac{u_i}{w_i}-1,-1,-1,-\frac{u_i^2}{2 w_i^2}+\frac{2 (1-v_i) u_i}{w_i^2}-\frac{v_i^2}{2 w_i^2}+\frac{2 v_i}{w_i^2}-\frac{3}{2 w_i^2},-\frac{u_i^2}{2 w_i^2}+\frac{2 (1-v_i) u_i}{w_i^2}-\frac{v_i^2}{2 w_i^2}+\frac{2 v_i}{w_i^2}-\frac{3}{2 w_i^2},-\frac{2 u_i v_i}{w_i^2}-\frac{2}{w_i},-\frac{2 u_i v_i}{w_i^2}-\frac{2}{w_i},\frac{4 u_i^2}{w_i^2}-\frac{6 (1-v_i) u_i}{w_i^2}+\frac{4 v_i^2}{w_i^2}-\frac{6 v_i}{w_i^2}+\frac{2}{w_i^2},-\frac{2 u_i}{w_i}-2,-\frac{2 (1-u_i)}{w_i},-2,6  \right)\end{dmath*}\begin{dmath*}
 \cRem{1}{0}{1}{1}{1}{0} = \left(0,0,0,0,1,0,\frac{1}{2},0,0,-3,2,\frac{1}{2},\frac{5}{2},-\frac{19}{2}  \right)\end{dmath*}\begin{dmath*}
 \cRem{1}{0}{1}{2}{0}{0} = \left(0,0,\frac{2 v_i}{w_i},0,0,\frac{v_i^2}{2 w_i^2},\frac{(1-u_i) v_i}{w_i^2}-\frac{(1-u_i){}^2}{2 w_i^2},\frac{v_i^2}{w_i^2},\frac{2 (1-u_i) v_i}{w_i^2}-\frac{(1-u_i){}^2}{w_i^2},\frac{v_i^2}{w_i^2},\frac{(1-u_i)}{w_i}-\frac{7}{2},-\frac{u_i}{w_i}-\frac{2 v_i}{w_i}+\frac{1}{(1-v_i) w_i},-\frac{3}{2},\frac{41}{8}  \right)\end{dmath*}\begin{dmath*}
 \cRem{2}{0}{0}{0}{0}{2} = \left(0,-\frac{u_i^2}{2 w_i^2},0,0,0,\frac{v_i u_i^2}{4 (1-u_i) w_i^2}-\frac{u_i^2}{8 w_i^2}-\frac{1}{8},\frac{u_i^2 v_i}{4 (1-u_i) w_i^2}-\frac{u_i^2}{8 w_i^2},\frac{u_i^2}{4 w_i^2}-\frac{1}{4},\frac{u_i^2 v_i}{2 (1-u_i) w_i^2}-\frac{u_i^2}{4 w_i^2},\frac{u_i^2 v_i}{2 (1-u_i) w_i^2}-\frac{u_i^2}{4 w_i^2},\frac{3 u_i^2}{4 (1-u_i) w_i}+\frac{1}{4},\frac{3 u_i}{4 w_i}-\frac{3}{8},-\frac{1}{8},\frac{15}{32}  \right)\end{dmath*}\begin{dmath*}
 \cRem{2}{0}{0}{0}{1}{1} = \left(0,\frac{u_i}{2 w_i},0,0,0,-\frac{(1-v_i)}{4 w_i},-\frac{u_i}{4 w_i},-\frac{(1-v_i)}{2 w_i},-\frac{u_i}{2 w_i},-\frac{u_i}{2 w_i},\frac{5}{8},-\frac{5}{4},-\frac{5}{8},\frac{19}{8}  \right)\end{dmath*}\begin{dmath*}
 \cRem{2}{0}{0}{0}{2}{0} = \left(0,0,0,\frac{1}{2},0,\frac{3}{8},0,0,0,-1,-\frac{1}{4},\frac{11}{8},\frac{9}{8},-\frac{15}{4}  \right)\end{dmath*}\begin{dmath*}
 \cRem{2}{0}{0}{1}{0}{1} = \left(0,\frac{(1-u_i) u_i}{w_i^2}-\frac{u_i v_i}{2 w_i^2},0,0,0,-\frac{u_i}{4 w_i}-\frac{1}{2},-\frac{u_i}{4 w_i},-\frac{u_i}{w_i}-1,-\frac{u_i}{2 w_i},-\frac{u_i}{2 w_i},1-\frac{u_i}{2 w_i},\frac{7 v_i}{4 w_i}-\frac{9 (1-u_i)}{4 w_i},-\frac{3}{4},\frac{11}{4}  \right)\end{dmath*}\begin{dmath*}
 \cRem{2}{0}{0}{1}{1}{0} = \left(0,0,0,1,0,\frac{1}{2},0,0,0,-\frac{5}{2},-\frac{3}{8},\frac{9}{4},\frac{15}{8},-\frac{31}{4}  \right)\end{dmath*}\begin{dmath*}
 \cRem{2}{0}{0}{2}{0}{0} = \left(1,\frac{(1-u_i) v_i}{w_i^2}-\frac{3 (1-u_i){}^2}{2 w_i^2},0,-\frac{3}{2},0,\frac{(1-u_i) v_i}{2 w_i^2}-\frac{7 (1-u_i){}^2}{8 w_i^2},\frac{(1-u_i) v_i}{2 w_i^2}-\frac{7 (1-u_i){}^2}{8 w_i^2},\frac{5 v_i^2}{4 w_i^2}-\frac{2 (1-u_i) v_i}{w_i^2},\frac{(1-u_i) v_i}{w_i^2}-\frac{7 (1-u_i){}^2}{4 w_i^2},\frac{5 (1-u_i){}^2}{4 w_i^2}-\frac{5 v_i (1-u_i)}{w_i^2}+\frac{3 v_i^2}{w_i^2},\frac{3 v_i}{2 w_i}-\frac{7 (1-u_i)}{4 w_i},-\frac{v_i^2}{4 (1-v_i) w_i},-\frac{7}{4},\frac{201}{32}  \right)\end{dmath*}\begin{dmath*}
 \cRem{2}{0}{1}{0}{0}{3} = \left(0,-\frac{3 u_i^2}{2 w_i^2},0,0,0,-\frac{u_i^3}{2 w_i^3}+\frac{3 v_i u_i^2}{4 (1-u_i) w_i^2}-\frac{9 u_i^2}{8 w_i^2}-\frac{1}{8},-\frac{5 u_i^4}{8 (1-u_i) w_i^3}+\frac{7 u_i^3}{4 (1-u_i) w_i^3}-\frac{3 v_i^2 u_i^2}{4 (1-u_i) w_i^3}+\frac{15 v_i u_i^2}{8 w_i^3}-\frac{9 u_i^2}{8 (1-u_i) w_i^3},-\frac{u_i^3}{4 w_i^3}-\frac{3 (1-v_i) u_i^2}{4 w_i^3}-\frac{1}{4},-\frac{5 u_i^4}{4 (1-u_i) w_i^3}+\frac{7 u_i^3}{2 (1-u_i) w_i^3}-\frac{3 v_i^2 u_i^2}{2 (1-u_i) w_i^3}+\frac{15 v_i u_i^2}{4 w_i^3}-\frac{9 u_i^2}{4 (1-u_i) w_i^3},-\frac{5 u_i^4}{4 (1-u_i) w_i^3}+\frac{7 u_i^3}{2 (1-u_i) w_i^3}-\frac{3 v_i^2 u_i^2}{2 (1-u_i) w_i^3}+\frac{15 v_i u_i^2}{4 w_i^3}-\frac{9 u_i^2}{4 (1-u_i) w_i^3},\frac{9 u_i^4}{4 (1-u_i){}^2 w_i^2}+\frac{5 v_i u_i^3}{4 (1-u_i){}^2 w_i^2}-\frac{3 u_i^3}{(1-u_i){}^2 w_i^2}+\frac{3 (1-v_i) u_i^2}{4 (1-u_i){}^2 w_i^2}+\frac{5}{24},-\frac{9 u_i^2}{4 w_i^2}+\frac{5 (1-v_i) u_i}{4 w_i^2}-\frac{11}{24},-\frac{1}{4},\frac{223}{288}-\frac{u_i^2}{2 (1-u_i) w_i}  \right)\end{dmath*}\begin{dmath*}
 \cRem{2}{0}{1}{0}{1}{2} = \left(0,\frac{u_i}{w_i},0,0,0,\frac{u_i^2}{4 w_i^2}-\frac{1}{4},\frac{u_i^2}{4 w_i^2},\frac{u_i^2}{2 w_i^2}-\frac{1}{2},\frac{u_i^2}{2 w_i^2},\frac{u_i^2}{2 w_i^2},\frac{u_i^2}{2 (1-u_i) w_i}+\frac{5}{8},\frac{u_i}{2 w_i}-\frac{3}{2},-\frac{7}{8},\frac{157}{48}  \right)\end{dmath*}\begin{dmath*}
 \cRem{2}{0}{1}{0}{2}{1} = \left(0,0,0,\frac{1}{2},0,\frac{3}{8},0,0,0,-\frac{3}{2},\frac{1}{8},\frac{11}{8},\frac{3}{2},-\frac{131}{24}  \right)\end{dmath*}\begin{dmath*}
 \cRem{2}{0}{1}{0}{3}{0} = \left(0,0,0,0,0,0,0,0,0,\frac{1}{2},-\frac{3}{8},-\frac{1}{4},-\frac{5}{8},\frac{157}{72}  \right)\end{dmath*}\begin{dmath*}
 \cRem{2}{0}{1}{1}{0}{2} = \left(0,\frac{2 (1-u_i) u_i}{w_i^2}-\frac{u_i v_i}{w_i^2},0,0,0,\frac{u_i^3}{2 w_i^3}+\frac{2 (1-u_i) (1-v_i) u_i}{w_i^3}-\frac{3 (1-v_i) v_i u_i}{2 w_i^3}-\frac{(1-v_i){}^3}{2 w_i^3},\frac{(1-u_i) u_i (1-v_i)}{2 w_i^3},-\frac{(1-v_i){}^3}{w_i^3}-\frac{2 u_i^2 (1-v_i)}{w_i^3}+\frac{3 u_i (1-v_i)}{w_i^3}-\frac{2 u_i v_i (1-v_i)}{w_i^3},\frac{(1-u_i) u_i (1-v_i)}{w_i^3},\frac{v_i u_i^2}{w_i^3}+\frac{u_i}{w_i^2},-\frac{v_i u_i^2}{2 (1-u_i) w_i^2}+\frac{3 u_i^2}{2 w_i^2}+1,\frac{3 (1-u_i) u_i}{2 w_i^2}-\frac{(1-u_i) (1-v_i)}{2 w_i^2}-\frac{9}{4},-\frac{5}{4},\frac{u_i}{2 w_i}+\frac{35}{8}  \right)\end{dmath*}\begin{dmath*}
 \cRem{2}{0}{1}{1}{1}{1} = \left(0,0,0,1,0,\frac{1}{2},0,0,0,-\frac{7}{2},\frac{1}{8},\frac{9}{4},\frac{19}{8},-\frac{21}{2}  \right)\end{dmath*}\begin{dmath*}
 \cRem{2}{0}{1}{1}{2}{0} = \left(0,0,0,0,0,0,0,0,0,1,-\frac{1}{2},-\frac{1}{2},-1,\frac{47}{12}  \right)\end{dmath*}\begin{dmath*}
 \cRem{2}{0}{1}{2}{0}{1} = \left(1,\frac{(1-u_i) v_i}{w_i^2}-\frac{3 (1-u_i){}^2}{2 w_i^2},-\frac{u_i}{w_i}-1,-\frac{3}{2},-1,\frac{v_i^3}{2 w_i^3}-\frac{5 (1-u_i) v_i^2}{2 w_i^3}+\frac{33 (1-u_i) v_i}{8 w_i^3}-\frac{29 (1-u_i) u_i v_i}{8 w_i^3}-\frac{17 (1-u_i){}^2}{8 w_i^3}+\frac{7 (1-u_i){}^2 u_i}{8 w_i^3},\frac{v_i^3}{2 w_i^3}-\frac{5 (1-u_i) v_i^2}{2 w_i^3}+\frac{33 (1-u_i) v_i}{8 w_i^3}-\frac{29 (1-u_i) u_i v_i}{8 w_i^3}-\frac{17 (1-u_i){}^2}{8 w_i^3}+\frac{7 (1-u_i){}^2 u_i}{8 w_i^3},-\frac{v_i^3}{4 w_i^3}-\frac{3 (1-u_i) v_i^2}{4 w_i^3}+\frac{7 (1-u_i) v_i}{2 w_i^3}-\frac{5 (1-u_i) u_i v_i}{2 w_i^3}-\frac{5 (1-u_i){}^2}{2 w_i^3},-\frac{3 (1-u_i) u_i^2}{4 w_i^3}+\frac{4 (1-u_i) u_i}{w_i^3}-\frac{17 (1-u_i) v_i u_i}{4 w_i^3}-\frac{2 (1-u_i) v_i^2}{w_i^3}-\frac{13 (1-u_i)}{4 w_i^3}+\frac{21 (1-u_i) v_i}{4 w_i^3},-\frac{11 v_i^3}{2 w_i^3}+\frac{29 (1-u_i) v_i^2}{2 w_i^3}-\frac{45 (1-u_i) v_i}{4 w_i^3}+\frac{49 (1-u_i) u_i v_i}{4 w_i^3}+\frac{9 (1-u_i){}^2}{4 w_i^3}-\frac{19 (1-u_i){}^2 u_i}{4 w_i^3},-\frac{u_i^2}{4 w_i^2}-\frac{9 v_i u_i}{2 w_i^2}+\frac{15 u_i}{4 w_i^2}-\frac{7 (1-v_i)}{2 w_i^2}+\frac{11 (1-v_i) v_i}{4 w_i^2},\frac{(1-u_i) v_i}{w_i^2}-\frac{5 (1-u_i){}^2}{2 w_i^2},-\frac{11}{4},\frac{257}{32}-\frac{3 v_i}{4 w_i}  \right)\end{dmath*}\begin{dmath*}
 \cRem{2}{0}{1}{2}{1}{0} = \left(0,0,0,0,1,0,\frac{1}{2},0,0,-\frac{7}{2},2,\frac{3}{4},\frac{11}{4},-\frac{32}{3}  \right)\end{dmath*}\begin{dmath*}
 \cRem{2}{0}{1}{3}{0}{0} = \left(0,0,\frac{3 v_i}{w_i},0,0,\frac{3 (1-u_i){}^2 v_i}{4 w_i^3}-\frac{v_i^3}{2 w_i^3},-\frac{(1-u_i){}^3}{2 w_i^3}+\frac{9 v_i (1-u_i){}^2}{4 w_i^3}-\frac{3 v_i^2 (1-u_i)}{2 w_i^3},\frac{3 (1-u_i){}^2 v_i}{2 w_i^3}-\frac{v_i^3}{w_i^3},-\frac{(1-u_i){}^3}{w_i^3}+\frac{9 v_i (1-u_i){}^2}{2 w_i^3}-\frac{3 v_i^2 (1-u_i)}{w_i^3},\frac{3 (1-u_i){}^2 v_i}{2 w_i^3}-\frac{v_i^3}{w_i^3},-\frac{17 (1-u_i){}^2}{6 w_i^2}+\frac{101 v_i (1-u_i)}{12 w_i^2}-\frac{61 v_i^2}{12 w_i^2},-\frac{10 v_i^3}{3 (1-v_i) w_i^2}+\frac{59 u_i v_i^3}{12 (1-v_i){}^2 w_i^2}+\frac{25 v_i^2}{4 (1-v_i) w_i^2}-\frac{19 u_i v_i^2}{2 (1-v_i){}^2 w_i^2}-\frac{13 v_i}{4 (1-v_i) w_i^2}+\frac{13 u_i v_i}{2 (1-v_i){}^2 w_i^2}+\frac{13 u_i^2}{12 w_i^2}+\frac{13}{12 (1-v_i) w_i^2}-\frac{13 u_i}{6 (1-v_i){}^2 w_i^2},-\frac{7}{4},\frac{115 v_i^2}{18 (1-v_i) w_i}-\frac{221 v_i}{18 (1-v_i) w_i}-\frac{221 u_i}{36 w_i}+\frac{221}{36 (1-v_i) w_i}  \right)\end{dmath*}\begin{dmath*}
 \cRem{3}{0}{0}{0}{0}{3} = \left(0,\frac{u_i^3}{3 w_i^3},0,0,0,\frac{v_i^2 u_i^3}{12 (1-u_i){}^2 w_i^3}-\frac{v_i u_i^3}{3 (1-u_i) w_i^3}+\frac{7 u_i^3}{36 w_i^3}-\frac{1}{18},\frac{v_i^2 u_i^3}{12 (1-u_i){}^2 w_i^3}-\frac{v_i u_i^3}{3 (1-u_i) w_i^3}+\frac{7 u_i^3}{36 w_i^3},-\frac{u_i^3}{9 w_i^3}-\frac{1}{9},\frac{v_i^2 u_i^3}{6 (1-u_i){}^2 w_i^3}-\frac{2 v_i u_i^3}{3 (1-u_i) w_i^3}+\frac{7 u_i^3}{18 w_i^3},\frac{v_i^2 u_i^3}{6 (1-u_i){}^2 w_i^3}-\frac{2 v_i u_i^3}{3 (1-u_i) w_i^3}+\frac{7 u_i^3}{18 w_i^3},\frac{5 v_i u_i^3}{36 (1-u_i){}^2 w_i^2}-\frac{7 u_i^3}{12 (1-u_i) w_i^2}+\frac{13}{108},-\frac{20 u_i^2}{27 w_i^2}+\frac{49 (1-v_i) u_i}{108 w_i^2}-\frac{17 (1-v_i){}^2}{108 w_i^2},-\frac{1}{27},\frac{109}{648}-\frac{11 u_i^2}{36 (1-u_i) w_i}  \right)\end{dmath*}\begin{dmath*}
 \cRem{3}{0}{0}{0}{1}{2} = \left(0,-\frac{u_i^2}{3 w_i^2},0,0,0,\frac{v_i u_i^2}{6 (1-u_i) w_i^2}-\frac{u_i^2}{12 w_i^2}-\frac{1}{12},\frac{u_i^2 v_i}{6 (1-u_i) w_i^2}-\frac{u_i^2}{12 w_i^2},\frac{u_i^2}{6 w_i^2}-\frac{1}{6},\frac{u_i^2 v_i}{3 (1-u_i) w_i^2}-\frac{u_i^2}{6 w_i^2},\frac{u_i^2 v_i}{3 (1-u_i) w_i^2}-\frac{u_i^2}{6 w_i^2},\frac{u_i^2}{2 (1-u_i) w_i}+\frac{7}{36},\frac{u_i}{2 w_i}-\frac{7}{24},-\frac{7}{72},\frac{55}{144}  \right)\end{dmath*}\begin{dmath*}
 \cRem{3}{0}{0}{0}{2}{1} = \left(0,\frac{u_i}{3 w_i},0,0,0,-\frac{(1-v_i)}{6 w_i},-\frac{u_i}{6 w_i},-\frac{(1-v_i)}{3 w_i},-\frac{u_i}{3 w_i},-\frac{u_i}{3 w_i},\frac{17}{36},-\frac{11}{12},-\frac{4}{9},\frac{131}{72}  \right)\end{dmath*}\begin{dmath*}
 \cRem{3}{0}{0}{0}{3}{0} = \left(0,0,0,\frac{1}{3},0,\frac{11}{36},0,0,0,-\frac{13}{18},-\frac{1}{3},\frac{241}{216},\frac{169}{216},-\frac{3487}{1296}  \right)\end{dmath*}\begin{dmath*}
 \cRem{3}{0}{0}{1}{0}{2} = \left(0,\frac{u_i^2 v_i}{6 w_i^3}-\frac{(1-u_i) u_i^2}{2 w_i^3},0,0,0,\frac{v_i u_i^2}{6 (1-u_i) w_i^2}-\frac{5 u_i^2}{24 w_i^2}-\frac{1}{8},\frac{u_i^2 v_i}{6 (1-u_i) w_i^2}-\frac{5 u_i^2}{24 w_i^2},\frac{u_i^2}{4 w_i^2}-\frac{1}{4},\frac{u_i^2 v_i}{3 (1-u_i) w_i^2}-\frac{5 u_i^2}{12 w_i^2},\frac{u_i^2 v_i}{3 (1-u_i) w_i^2}-\frac{5 u_i^2}{12 w_i^2},-\frac{v_i u_i^2}{2 (1-u_i) w_i^2}+\frac{5 u_i^2}{6 w_i^2}+\frac{1}{4},-\frac{7 u_i^2}{6 w_i^2}-\frac{7 v_i u_i}{6 w_i^2}+\frac{19 u_i}{12 w_i^2}-\frac{v_i^2}{3 w_i^2}+\frac{3 v_i}{4 w_i^2}-\frac{5}{12 w_i^2},-\frac{1}{12},\frac{u_i}{4 w_i}+\frac{103}{288}  \right)\end{dmath*}\begin{dmath*}
 \cRem{3}{0}{0}{1}{1}{1} = \left(0,\frac{(1-u_i) u_i}{2 w_i^2}-\frac{u_i v_i}{3 w_i^2},0,0,0,-\frac{u_i}{6 w_i}-\frac{1}{4},-\frac{u_i}{6 w_i},-\frac{(1-v_i)}{2 w_i},-\frac{u_i}{3 w_i},-\frac{u_i}{3 w_i},\frac{5}{8}-\frac{u_i}{6 w_i},\frac{7 v_i}{6 w_i}-\frac{4 (1-u_i)}{3 w_i},-\frac{13}{24},\frac{19}{9}  \right)\end{dmath*}\begin{dmath*}
 \cRem{3}{0}{0}{1}{2}{0} = \left(0,0,0,\frac{1}{2},0,\frac{3}{8},0,0,0,-\frac{7}{6},-\frac{4}{9},\frac{3}{2},\frac{19}{18},-\frac{71}{18}  \right)\end{dmath*}\begin{dmath*}
 \cRem{3}{0}{0}{2}{0}{1} = \left(0,\frac{u_i (1-u_i){}^2}{w_i^3}-\frac{u_i v_i (1-u_i)}{w_i^3}+\frac{u_i v_i^2}{3 w_i^3},0,0,0,\frac{u_i (1-u_i){}^2}{12 w_i^3}-\frac{u_i}{6 w_i}-\frac{1}{2},-\frac{u_i (1-u_i){}^2}{12 w_i^3}+\frac{u_i v_i (1-u_i)}{3 w_i^3}-\frac{u_i v_i^2}{6 w_i^3},-\frac{(1-v_i){}^3}{w_i^3}-\frac{u_i^2 (1-v_i)}{w_i^3}+\frac{13 u_i v_i^2}{6 w_i^3}+\frac{2 u_i}{w_i^3}-\frac{4 u_i v_i}{w_i^3},-\frac{u_i (1-u_i){}^2}{6 w_i^3}+\frac{2 u_i v_i (1-u_i)}{3 w_i^3}-\frac{u_i v_i^2}{3 w_i^3},-\frac{u_i (1-u_i){}^2}{6 w_i^3}+\frac{2 u_i v_i (1-u_i)}{3 w_i^3}-\frac{u_i v_i^2}{3 w_i^3},-\frac{5 (1-u_i) u_i}{6 w_i^2}+\frac{2 v_i u_i}{3 w_i^2}+1,-\frac{29 (1-u_i){}^2}{12 w_i^2}+\frac{23 v_i (1-u_i)}{6 w_i^2}-\frac{19 v_i^2}{12 w_i^2},-\frac{7}{12},\frac{83}{36}-\frac{(1-u_i)}{6 w_i}  \right)\end{dmath*}\begin{dmath*}
 \cRem{3}{0}{0}{2}{1}{0} = \left(0,0,0,1,0,\frac{1}{2},0,0,0,-\frac{17}{6},-\frac{23}{36},\frac{29}{12},\frac{16}{9},-\frac{295}{36}  \right)\end{dmath*}\begin{dmath*}
 \cRem{3}{0}{0}{3}{0}{0} = \left(1,-\frac{11 (1-u_i){}^3}{6 w_i^3}+\frac{5 v_i (1-u_i){}^2}{2 w_i^3}-\frac{v_i^2 (1-u_i)}{w_i^3},0,-\frac{11}{6},0,-\frac{85 (1-u_i){}^3}{72 w_i^3}+\frac{11 v_i (1-u_i){}^2}{8 w_i^3}-\frac{v_i^2 (1-u_i)}{2 w_i^3},-\frac{85 (1-u_i){}^3}{72 w_i^3}+\frac{11 v_i (1-u_i){}^2}{8 w_i^3}-\frac{v_i^2 (1-u_i)}{2 w_i^3},-\frac{49 v_i^3}{36 w_i^3}+\frac{15 (1-u_i) v_i^2}{4 w_i^3}-\frac{3 (1-u_i){}^2 v_i}{w_i^3},-\frac{85 (1-u_i){}^3}{36 w_i^3}+\frac{11 v_i (1-u_i){}^2}{4 w_i^3}-\frac{v_i^2 (1-u_i)}{w_i^3},\frac{49 (1-u_i){}^3}{36 w_i^3}-\frac{101 v_i (1-u_i){}^2}{12 w_i^3}+\frac{61 v_i^2 (1-u_i)}{6 w_i^3}-\frac{67 v_i^3}{18 w_i^3},-\frac{247 (1-u_i){}^2}{108 w_i^2}+\frac{419 v_i (1-u_i)}{108 w_i^2}-\frac{101 v_i^2}{54 w_i^2},\frac{5 v_i^3}{12 (1-v_i) w_i^2}-\frac{25 u_i v_i^3}{36 (1-v_i){}^2 w_i^2}-\frac{3 v_i^2}{4 (1-v_i) w_i^2}+\frac{3 u_i v_i^2}{4 (1-v_i){}^2 w_i^2},-\frac{247}{108},\frac{21811 v_i^2}{2592 (1-v_i) w_i}+\frac{21955 u_i v_i}{2592 (1-v_i) w_i}-\frac{21955 v_i}{1296 (1-v_i) w_i}+\frac{21955 (1-u_i)}{2592 (1-v_i) w_i}  \right)\end{dmath*}
 \end{dgroup*}

\bibliographystyle{utcaps}
\bibliography{ThesisINSPIRE}

\end{fmffile}
\end{document}